\documentclass{aa}
\usepackage[varg]{txfonts}
\usepackage{subfig}
\usepackage{esvect}
\usepackage{natbib}
\usepackage{overpic}
\usepackage{hyperref}
\usepackage{comment}
\usepackage{ulem}
\usepackage{adjustbox}

\def\i{\,{\sc i}}
\def\ii{\,{\sc ii}}
\def\iii{\,{\sc iii}}
\def\Ms{$\textrm{M}_{\odot}$}

\def\nb{\textsc{nbursts}}

\newcommand{\Ha}{H$\alpha$}
\newcommand{\Hb}{H$\beta$}
\newcommand{\Hg}{H$\gamma$}

\usepackage{tikz}
\usetikzlibrary{arrows}
\usetikzlibrary{arrows.meta}
\usetikzlibrary{3d}
\usepackage{xparse}
\usetikzlibrary{math} 
\usetikzlibrary{shapes.misc} 

\tikzset{
dot/.style = {circle, fill, minimum size=#1,
              inner sep=0pt, outer sep=0pt},
dot/.default = 6pt 
}

\begin{document}
\title{A group of merging galaxies falling onto Abell 2142}
\author{Aashiya Anitha Shaji\inst{1}
            \and
          Anaëlle Hallé\inst{1}
          \and
          Damir Gasymov\inst{2,3}
          \and
           Anne-Laure Melchior\inst{1}
        \and
           Fran\c coise Combes\inst{1,4}        
          \and 
          Andrea Cattaneo\inst{1}
          }

\institute{LUX, Observatoire de Paris, Sorbonne Universit\'{e}, Universit\'{e} PSL, CNRS, F-75014, Paris, France\\  
\email{Aashiya.Shaji@obspm.fr} 
        \and
        Astronomisches Rechen-Institut, Zentrum für Astronomie der Universität Heidelberg, Mönchhofstr. 12–14, 69120 Heidelberg, Germany
    \and
Sternberg Astronomical Institute, Moscow M.V. Lomonosov State University, Universitetskij pr., 13,  Moscow, 119234, Russia
          \and
Coll\`{e}ge de France, 11 Place Marcelin Berthelot, 75005 Paris, France}

\abstract{
Galaxy clusters produce a very hostile environment to galaxies, whose gas gets stripped by ram-pressure, suffer galaxy interactions and witness quenching of their star formation. Clusters continue their growth not only through galaxy accretion but also through galaxy group infall, such as Abell 2142, directly connected to the cosmic web. Our goal is to study the physical and dynamical state of the most conspicuous infalling group, on a filament projected at 1.3 Mpc from the Abell 2142 center. The galaxy group is the leading edge of a spectacular trailing X-ray tail, 700 kpc in length, of hot gas stripped by ram-pressure. The infalling galaxies are not yet quenched, and they are ideal objects to study the transformation processes due to the cluster environment. We use integral field spectroscopy from MaNGA to derive stellar and gas kinematics, and MegaCam for ugr photometry. Stellar populations (with age and metallicity)  are obtained through full-spectrum fitting using Nburst. The gas kinematics and excitation are derived from the line emission of H$\alpha$, [NII], [OIII] and H$\beta$. The group contains four galaxies, of which two are merging, and partly superposing on the line of sight. With a simple parametric model for each velocity field, we succeed in disentangling the contribution of each galaxy, and derive their physical state and kinematics. The galaxies are perturbed and intra-group gas is observed  as tidal tails and loops. They are mainly disks in rotation, although some regions reveal elevated dispersion, typical of out of equilibrium gas. All galaxies show sustained star formation, with a global star formation rate of 45 M$_\odot$/yr. We conclude that the long X-ray tail must have come from the hot intra-group medium, present before the group infall, and does not correspond to the ram-pressure stripping of the galaxy gas. The galaxy interactions and merging within the group are still enhancing the star formation, from the galaxy disks, which are still rich in dense gas.
}
\keywords{Galaxies: star formation --
       Galaxies: clusters: general
       Galaxies: groups: general
 	 Galaxies: interactions --
 		Techniques: imaging spectroscopy}
\maketitle

\section{Introduction}
\label{sect:intro}
Galaxy clusters are the largest gravitationally bound structures in the universe, formed through the nonlinear growth of primordial density fluctuations \citep{2014Vikhlinin}. These massive assemblies of galaxies, dominated by dark matter and hot gas visible through X-ray emission \citep{1977Bahcall_review, 2019Bykov}, evolve slowly and retain important clues about the universe's formation through the ongoing accretion of galaxies \citep[e.g.][]{2006Springel}. Galaxies within clusters tend to be redder, more compact, and have larger Sérsic indices compared to their counterparts in the field \citep{2017Pranger}. They also experience gas depletion and reduced star formation rates, a result of the dense cluster environment \citep{2006Boselli&Gavazzi, 2004Bower&Balogh}. It is well established that environment significantly influences galaxy evolution, with galaxies in high-density regions undergoing markedly different evolutionary paths than those in lower-density environments \citep{2015Pasquali, 2007Mateus}.

One of the key processes affecting galaxies as they fall into a cluster is ram pressure stripping (RPS). As galaxies move through the hot intracluster medium (ICM), the pressure exerted by the ICM can strip away their gas, leaving the stars largely unaffected \citep{1972Gunn,2016Fillingham,2016Steinhauser}. Galaxies with RPS tails, commonly known as jellyfish galaxies, exhibit this gas removal in various phases—neutral hydrogen (HI) \citep{2007Chung, 2012Scott}, molecular gas (such as CO) \citep{2013Jachym,2014Jachym,2017Jachym}, and even X-ray observations \citep{2006Sun}. The process removes gas in layers, akin to peeling onion skins, with the hot diffuse gas being stripped very efficiently, while molecular gas remains more resistant \citep{2009Bekki,2019Hausammann}. In high-density environments, RPS can significantly deplete the interstellar medium (ISM) of the galaxy, suppressing star formation \citep[and references therein]{2022Boselli_review}. While enhanced star formation in the stripped tails is expected as gas is removed from the disk, what's surprising is the temporary boost in star formation within the disk itself, as the increased pressure compresses the ISM gas into stars before stripping begins \citep{2018Vulcani}.

While individual galaxies face significant transformation as they move through the cluster environment, many galaxies do not fall into clusters alone. Recent simulations by \citet{2022Kuchner} suggest that about 12\% of the galaxies that fall onto clusters arrive in groups, which are smaller gravitationally bound systems containing a few to dozens of galaxies. When galaxy groups fall into a cluster, the infall triggers complex interactions, called "pre-processing" \citep{2018Bianconi}, including gravitational forces, enhanced gas stripping \citep{2021Kleiner} and an increased likelihood of mergers between galaxies \citep{2013Vijayaraghavan, 2020Benavides}. These mergers can significantly influence star formation, sometimes triggering bursts of activity \citep{1988Sanders, 1999Bekki, 2008Li} or, conversely, leading to quenching. 

Both galaxy mergers \citep{2024Comerford} and RPS are known to trigger active-galactic nuclei (AGN). \citet{2022Peluso} found a significantly higher AGN fraction in ram-pressure-stripped galaxies, especially in massive ones, aligning with \citet{2020Ricarte}, who noted that RPS enhances black hole accretion in massive galaxies during pericentric passage but suppresses it in less massive ones. \citet{2017Poggianti} reported a high incidence of AGN among extreme jellyfish galaxies, suggesting that ram pressure funnels gas toward galactic centers to trigger AGN activity, while \citet{2018Marshall} developed a model indicating that this triggering occurs within a specific range of ram pressures (P$_\text{ram} = 10^{-14} -10^{-13}$ Pa).

\begin{table}[h]
    \centering
    \begin{tabular}{|c|c|}
         \hline
         \textit{Properties} & \textit{Values}\\
         \hline
         Right Ascension$^1$ & 15$^\text{h}$58$^\text{m}$20$^\text{s}$\\
         Declination$^1$ & 27$^\circ14'00''$\\
         Redshift$^1$ & 0.0894\\
         $R_{200}$$^2$ & $2160 \pm 80$ kpc\\
         $M_{200}$$^2$ & $(12.5 \pm 1.3) \times 10^{14} M_\odot$ \\
         $R_{500}$$^3$& $1408 \pm 70$ kpc \\
         $M_{500}$$^3$& $(8.66 \pm 0.2) \times 10^{14} M_\odot$\\
         \hline
    \end{tabular}
    \caption{Properties of Abell 2142. References: $^1$ \citet[Planck Collaboration XXVII]{2016Planck}, $^2$\citet{2014Munari}, $^3$\citet{2016Tchernin}}
    \label{tab:A2142_properties}
\end{table}


Such group-cluster interactions are particularly evident in systems like Abell 2142, at $z=0.0894$ \citep{2018Bilton}, where the infall of galaxy groups drives much of the cluster's ongoing evolution. Its properties are summarised in Table \ref{tab:A2142_properties}. Former studies, such as those by \citet{1996Buote} utilizing ROSAT PSPC imagery, argue that A2142 has reached an advanced stage of a merger, a notion further explored by \citet{2000Markevitch} who identified it as a merger involving two subclusters using \textit{Chandra}. X-ray observations, such as those by \citet{2008Okabe}, reveal a complex core region marked by two cold fronts.
\citet{2018Liu} spectroscopically identified 868 galaxy members within 3.5 Mpc, unveiling substructures indicative of ongoing merging activity. This galactic ensemble serves as an ideal testing ground for numerical simulations and dark matter assembly \citep[e.g.][]{2014Munari,2016Munari}. Nested within a collapsing super-cluster \citep{2015Einasto,2020Einasto}, A2142 hosts detached long filaments that seamlessly connect to the cosmic web \citep[see also][]{2019Poopakun}. Recently, with the advent of LOFAR radio data, \citet{2023Bruno} unveiled a three-component giant radio halo within A2142, accompanied by numerous synchrotron relics tracing past collisions and related shocks.

Numerous groups are falling onto A2142 \citep{2018Einasto}. SDSS J155850.44+272323.9 (at $z = 0.0935$) is the main galaxy of a gas-rich infalling galaxy group studied in X-ray by \citet{2014Eckert,2017Eckert}. This galaxy group, located at around 1.3 Mpc (R$_{gg}$) northeast of the centre of Abell 2142, has been caught before its quenching and exhibits the largest star formation activity \citep[40\,$M_\odot$/yr,][]{2016Salim} of the whole cluster. It is located at the tip of a spectacular 700 kpc long X-ray tail typical of hot gas stripped by RPS. From the XMM-Newton observations, \citet{2014Eckert} estimated that this group (with a mass of a few $10^{13} M_{\odot}$) is falling onto the cluster with a relative velocity of about 1200 km/s, and while 95\,\% of the hot gas is located in the tail that was expelled from the group $\ge 600$ Myr ago, only 5\,\% remains in the galaxy group. The deep \textit{Chandra} observations of the group revealed significant AGN activity in two of the galaxies (marked in Figure \ref{fig:SDSS} as Galaxy a and Galaxy b) with X-ray luminosities $(2.4\pm0.3) \times 10^{42}$ erg/s and $(3.0\pm0.4)\times10^{42}$ erg/s, respectively, in the $[2-10]$ keV band \citep{2017Eckert}. They also identified a complex morphology in the tail characterized by an abrupt flaring beyond 250 kpc and a drop in metallicity beyond 300 kpc. Their study highlighted the role of magnetic draping and slow gas interactions with the ICM in shaping the tail, with the structure showing signs of turbulence and suppressed thermal conduction.

In this paper, we explore the kinematics of this complex system with the SDSS MaNGA data. In Sect. \ref{sect:data}, we describe the data. In Sect. \ref{sect:stellar}, we perform the stellar analysis while in Sect. \ref{sect:gas} we analyse the gas in the system. In Sect. \ref{sect:disc}, we discuss our results, and draw our conclusions in Sect. \ref{sect:conclu}. 
Throughout the paper, we assume a $\Lambda$CDM cosmology with $\Omega_m = 0.3$, $\Omega_\Lambda = 0.7$ and  $H_0=70$\,km\,s$^{-1}$/Mpc. 







\section{Data description and characteristics}
\label{sect:data}

\subsection{Optical IFU Data from MaNGA }
\label{ssect:MaNGA}
This galaxy group (MaNGA ID: 1-265404, plateifu = 9094-1902) was observed as part of the Mapping Nearby Galaxies at Apache Point Observatory (MaNGA) survey \citep{2015Bundy} on 17 April 2018. The spectra were acquired through the MaNGA instrument \citep{2015ADrory} mounted on the Sloan Foundation 2.5-meter Telescope, which is located at the Apache Point Observatory in New Mexico, USA. The MaNGA Data Analysis Pipeline (DAP) utilises the MILESHC template \citep{2019Westfall} for the stellar kinematics and MaSTAR stellar library to obtain the emission lines. The data products include a 3-D reduced data cube, a model data cube and 2-D maps of emission-line properties.  

The centre of the data cubes is located at RA = $15^h 58^m 50.4^s$, DEC = $27^{\circ} 23' 23.9''$, which does not necessarily coincide with the centre of the galaxy group. The dimensions of the data cubes are $32\times 32\times 4563$, where the former two values correspond to the spatial dimensions and the latter to the wavelength dimension. The spectral data is stored in the 'FLUX' extension, with each pixel containing information about the specific intensity in units of $1 \times 10^{-17}$ erg/s/cm$^2$/\AA/spaxel. The spectral cube covers a broad wavelength range, from 3600–10300 \AA\, at $R \sim 2000$. The pixel size is 0.5 arcseconds which corresponds to 0.87~kpc at $z=0.0935 $. The FWHM of the reconstructed Gaussian PSF in the $g$ or $r$ bands is of 2.6 arcseconds (4.5~kpc) and the FWHM spectral resolution ranges from $250$ km/s on the blue end to $150$ km/s on the red end. 

\subsection{MegaCam/MegaPipe ugr photometry}
\label{ssect:archival}

\begin{figure*}[h]
    \centering
    \includegraphics[scale = 0.5, keepaspectratio]{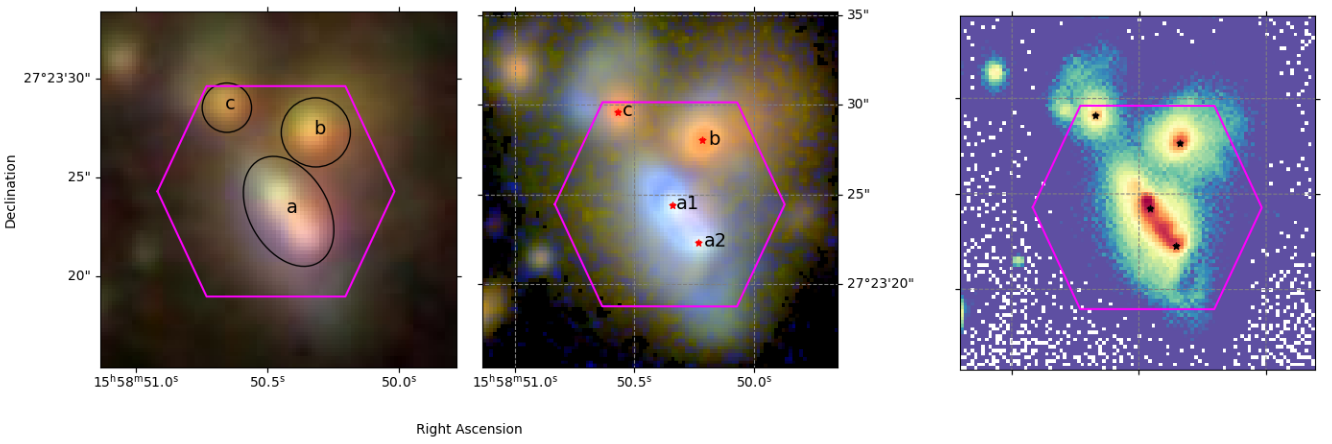} 
    \caption{SDSS $gri$ image (left), MegaPipe $rgu$ image (centre) and unsharp masking of the Megapipe $r$-band image (right), with the MaNGA field of view overlaid in magenta. The three main stellar regions are segregated and marked as ($a$,$b$,$c$) on the SDSS image. The centres of the regions, with a subdivision of the $a$ region into two galaxies, are shown on the MegaPipe images.}
    \label{fig:SDSS}
    
\end{figure*}

\begin{figure*}[h]
    \includegraphics[width=\textwidth]{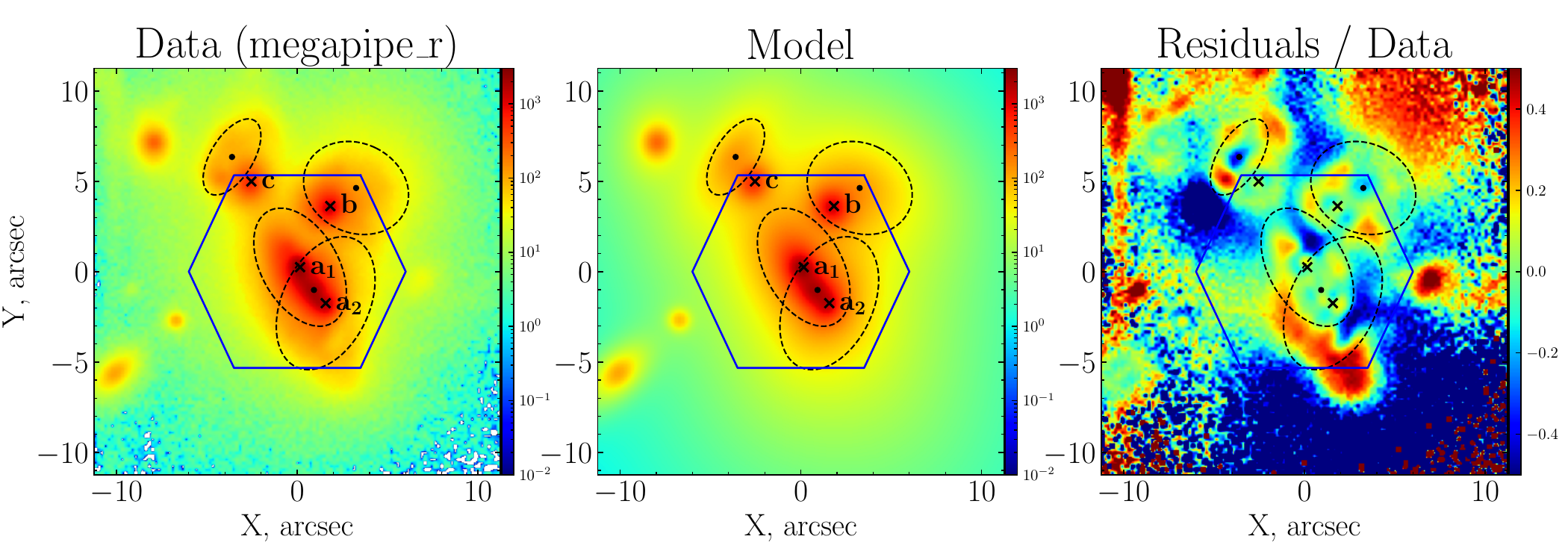}
    \caption{The photometric modelling of the CFHT MegaPipe $r$ band image using \textsc{galfit}. Blue hexagons show MaNGA field of view, ``x'''s mean the centers of the ``core'' components, dots near to \textit{c} and \textit{b} components mean the centers of the ``outskirt'' components, and dot between $a_1$ and $a_2$ is the center of ``bridge'' between them (see~Subsec.~\ref{subsec:St_pop_method}). Dashed ellipses show "extended" components of galaxies based on the parameters from the Table~\ref{tab:losvd_phot}. Third panel shows residuals-to-data ratio.}
    \label{fig:galfit_megapipe}
\end{figure*}

We used archival data from MegaCam, a state-of-the-art wide-field imaging facility at the Canada-France-Hawaii Telescope (CFHT). MegaCam is known for its 0.187-arcsecond resolution per pixel and accurate photometric calibrations up to 0.03 mag. The data from the MegaCam Image Stacking Pipeline \citep[MegaPipe;] []{2008Gwyn}, utilizing first-generation broadband filters, ugr, were used to create a combined RGB image (Fig. \ref{fig:SDSS}), photometric modelling (Fig. \ref{fig:galfit_megapipe}), and to perform unsharp masking (Fig. \ref{fig:SDSS}).
\begin{figure*}  
\centering
   \includegraphics[trim=0mm 0mm 0mm 10mm, clip, width=0.95\textwidth,keepaspectratio]{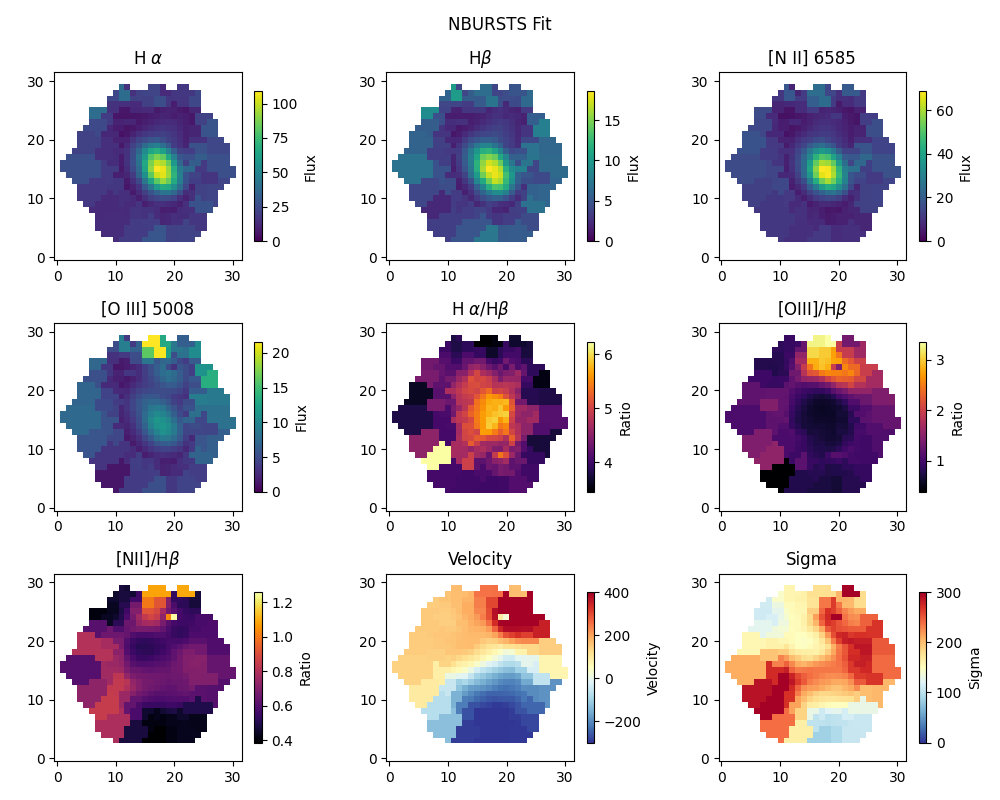}
\caption{Emission line maps from NBURST single gaussian fits. The main emission lines (H$\alpha$, H$\beta$, [NII]\,6585 and [OIII]\,5008) are displayed in arbitrary unit, as well as the line ratios, the velocity and velocity dispersion in km/s.}
    \label{fig:Nburst}
\end{figure*}
At first glance, we observe three main stellar structures in the left panel of Fig. \ref{fig:SDSS}. However, upon closer inspection, we notice two distinct colours in the MaNGA image of the central galaxy: the northern part of the galaxy (a) appears bluer than its southern part. To investigate this further, we applied a Gaussian filter with a standard deviation of $\sigma=0.75$ and generated a Lupton RGB image \citep[right panel of Fig. \ref{fig:SDSS}]{2004Lupton} using the Megapipe cutouts of the $r$-, $g$-, and $u$-bands as the red, green, and blue channels, respectively. This figure depicts that galaxies (b) and (c) are redder than the central galaxy, which can be attributed to either the older stellar population or the presence of dust. Galaxy (c) seems to be merging with a bluer galaxy which is, however, situated outside of the MaNGA field-of-view. 

To investigate potential substructures within the central galaxy, we applied a Gaussian convolution with a standard deviation of 0.5 to the original image, using the Gaussian2DKernel from Astropy. The resulting convolved image was then subtracted from the original data, scaled appropriately, and adjusted for dynamic range. Two distinct features (henceforth referred to as a1 and a2) become evident in the central region, revealing two galaxies (in Figure \ref{fig:SDSS}) instead of the one previously perceived. Moreover, the tidal tails and their directions are enhanced in this image.

We performed a \textsc{galfit} analysis \citep{Peng2002AJ....124..266P, Peng2010AJ....139.2097P} using the Megapipe r-band image.
It was the reddest band with the best seeing for correct decomposition and obtaining stellar luminosity.
The merging system is complex and cannot be easily modeled with symmetric and stationary components, as it contains star-forming regions, dust lanes, tidal tails, and shells. In our analysis, we assumed that each galaxy consists of a central ``core,'' while galaxies \textit{b} and \textit{c} also have extended ``outskirts'' components. The ``core'' components were modeled using a S\'ersic profile, while the extended components were fitted with an exponential disk profile. For galaxies $a_1$ and $a_2$, we added a compact exponential disk that acts as a ``bridge'' between them and provides a good fit. Galaxies $b$ and $c$ exhibited off-centered components, possibly due to ram pressure or other external factors. The final parameters of the galaxies are provided in Table~\ref{tab:losvd_phot}, and the image, best-fit model, and residuals are shown in Fig.~\ref{fig:galfit_megapipe}.

\subsection{Subaru photometry}
Subaru images in the $g$ and $r$ band are shown in the Appendix~\ref{app:subaru}. While some central pixels are blank, the images provide an even better view of structures such as tidal tails and bridges.

\section{Stellar analysis and first glance at the gas component}
\label{sect:stellar}
\subsection{Stellar populations}
One of our objectives was to estimate the stellar masses of the system’s components. Using \textsc{galfit}, we were able to decompose the photometric information of the system into its individual components and determine their luminosities. However, to derive the stellar mass, we require the mass-to-light ratio (M/L), which depends on the properties of the stellar populations. These properties can be determined using a full-spectrum fitting with a two-component technique. Before applying this method, several preliminary steps are necessary.

\subsubsection{One-component analysis}
The first step involves performing a full-spectrum fitting using a one-component model with \nb\ \citep{nburst_a, nburst_b}, in order to estimate the average properties of the stellar populations, including the age (T$_\mathrm{SSP}$) and metallicity ([Z/H]$_\mathrm{SSP}$), as well as the stellar and gas kinematics, such as velocity ($V$) and velocity dispersion ($\sigma$). For this analysis, we employed \nb\ with a grid of simple stellar population (SSP) models from the \textsc{e-miles} library \citep{Vazdekis2016MNRAS.463.3409V}.

\begin{figure}
    \centering
    \includegraphics[width=0.9\linewidth]{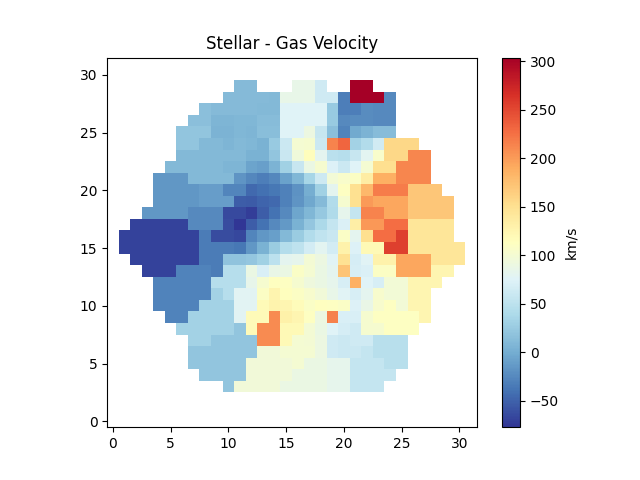}
    \caption{Difference between the stellar and gas velocity  }
    \label{fig:St-Gas}
\end{figure}

During each model evaluation in the $\chi^2$-minimization process (Eq.~\ref{eq:chi_sq}), the stellar population spectrum is interpolated from the model grid based on the given age (T$_\mathrm{SSP}$) and metallicity ([Z/H]$_\mathrm{SSP}$) values. The interpolated spectrum is then convolved with the stellar line-of-sight velocity distribution (LOSVD), which in our case is parameterized by a pure Gaussian. Additionally, the spectrum is multiplied by a polynomial continuum to account for the difference between the model and observed spectral shapes, which arises from imperfect spectral sensitivity calibration and the effects of dust extinction. For this, we employed multiplicative Legendre polynomials of the 19th degree (1 degree per each $150-200$ \AA\ of wavelength range and an odd degree is required). The model also includes a set of strong emission lines (\Hg, \Hb, [O\iii], [O\i], [N\ii], \Ha, [S\ii], etc.) possessing the same Gaussian kinematics but are treated independently from the stellar kinematics. These emission lines are additive components whose weight, i.e. emission line fluxes, are determined as part of a linear problem solved at each iteration of the non-linear minimization loop. Prior to the main minimization process, both the emission line templates and the grid of stellar population templates were pre-convolved with the wavelength-dependent instrumental resolution. The general formulas of \nb\ are:
\begin{equation}
    F^{model} = P_{n^{mult}} \cdot \sum_{N_{st}} p^{st} \cdot T^{st} \otimes \mathcal{L}^{st} + \sum_{N_{em}} p^{em} \cdot T^{em} \otimes \mathcal{L}^{em},
\end{equation}
\begin{equation}
    \chi^2 = \sum_{N_\lambda} p_\lambda (F_\lambda - F_\lambda^\mathrm{model})^2 / \Delta F_\lambda^2.
    \label{eq:chi_sq}
\end{equation}
Here, $P_{n^{mult}}$ denotes the multiplicative continuum in the form of a Legendre polynomial, while $N_{st}$ and $N_{em}$ represent the number of stellar components and emission lines, respectively. The terms $p^{st}$ and $p^{em}$ refer to the contributions of the stellar component (ranging from 0 to 1, $\sum_{N_{st}} p^{st} = 1$) and the emission line (the flux of the line). The symbols $T^{st}$ and $T^{em}$ denote models for the stellar and emission line components, and $\mathcal{L}^{st}$ and $\mathcal{L}^{em}$ represent the convolved parameterized LOSVDs with the spectrograph's line spread function (LSF). In Eq.~\ref{eq:chi_sq}, $p_{\lambda}$ is the mask for good wavelengths, $F_{\lambda}$ represents the observed spectrum, and $\Delta F_{\lambda}$ is the error associated with the spectrum.
The main results of this analysis are shown on Fig.~\ref{fig:Nburst} and Fig.~\ref{fig:St-Gas}. With this procedure, we confirmed the stellar and gas kinematic maps provided by MaNGA DAP \citep{2019Westfall}, and we obtained the stellar population parameters T$_\mathrm{SSP}$ and [Z/H]$_\mathrm{SSP}$, as well as the stellar spectra unconvolved with a parametric gaussian LOSVD, in each spatial bin. 

From Fig \ref{fig:Nburst}, we can draw several conclusions regarding the star-forming regions and the galaxies involved. The high H$\alpha$/H$\beta$ ratio observed at the centre suggests significant dust emission, indicating a rich star formation activity. Additionally, Galaxy (b) appears to be a strong active galactic nucleus (AGN), while Galaxy (a1) seems to be obscured by dust, impacting its AGN visibility. Furthermore, the high-velocity dispersion in the central region aligns with expectations; however, the anomalous dispersions noted in the bottom left and top right areas may be indicative of some unknown perturbations affecting the dynamics within this system. A velocity difference of 300 km/s between the gas and stellar components seen in Fig \ref{fig:St-Gas}, combined with a high star formation rate and the presence of an AGN, may indicate that outflows from the AGN and feedback from star formation processes are driving the observed kinematic features.

\subsubsection{Two-component analysis}
\label{subsec:St_pop_method}
To investigate the two-component stellar kinematics in our sample and perform full-spectrum fitting with two stellar components, an accurate initial approximation of the kinematics of the components was required and we employed a non-parametric LOSVD recovery approach, successfully used in previous studies \citep[e.g.,][]{Katkov2016MNRAS.461.2068K,Kasparova2020MNRAS.493.5464K,Katkov2024ApJ...962...27K}, with a detailed description provided in \citep{Gasymov2024ASPC..535..279G}. The key concept of this method is to find a convolution kernel such that, when convolved with the original stellar spectrum, the result best reproduces the observed spectra, see Fig.~\ref{fig:2comp_fit}.
In the previous step, this kernel was parameterized as a pure Gaussian, but now we solve an ill-posed linear problem to recover its complex shape. This problem is addressed using a least-squares method, with additional L2 norm minimization and “edge” regularization to smooth the solution and suppress spurious high-velocity peaks due to the limited signal-to-noise ratio.

\begin{figure}[h]
    \includegraphics[width=0.5\textwidth]{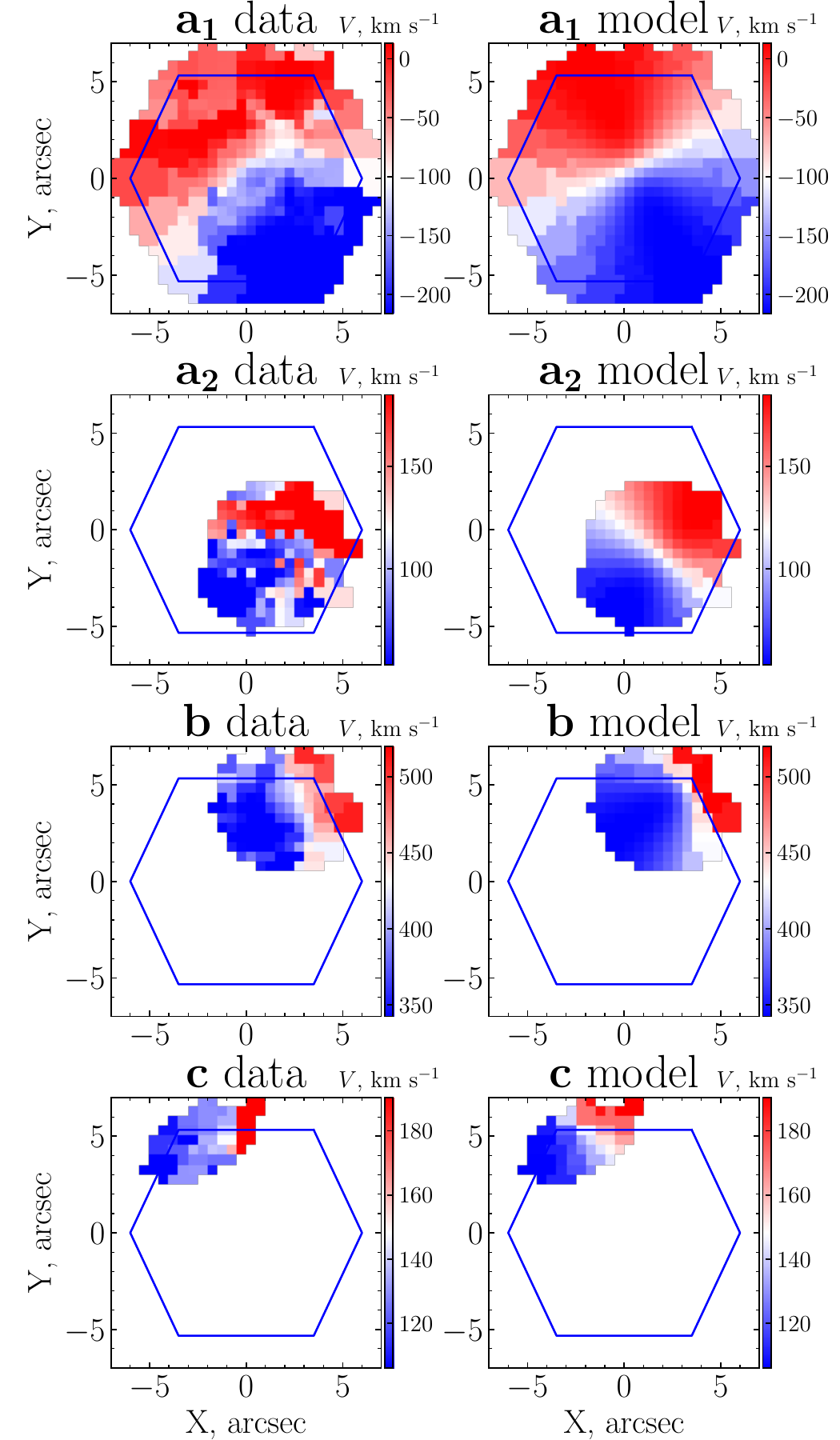}
    \caption{The kinematical modelling of the stellar recovered LOSVD. The panels show the velocity maps of the different components and their corresponding models (see Subsection~\ref{subsec:St_pop_method}). Blue hexagons show the field of view covered by the MaNGA observations.}
    \label{fig:LOSVD_models}
\end{figure}

As displayed Fig.~\ref{fig:LOSVD_bins}, the solution is a vector of the LOSVD for each spatial bin, allowing us to trace the three-dimensional kinematics of the system (in X-Y-V coordinates). In particular, the spaxels in the “merging” regions reveal prominent two- or three-component structures, providing clear evidence of distinct kinematic behavior in these galaxies, which cannot be accurately described by a single Gaussian parametric LOSVD. This multi-component structure now enables us to model the relative systemic velocities between the galaxies.

We selected the regions corresponding to components a$_2$, b, and c (Fig.~\ref{fig:LOSVD_models} and \ref{fig:LOSVD_bins}) and assumed Gaussian kinematics for these components within the designated circular regions, defined in Fig.~\ref{fig:LOSVD_bins}. Using one-, two-, and three-Gaussian models in corresponding spaxels, we then examined the kinematic maps of each galaxy. These maps exhibit a consistent rotation pattern, which we modeled using a circular rotation model:
\begin{equation}
V(r, i, \theta) = V_0 (r) \sin i  \cos \theta + V_\mathrm{sys},
\label{eq:V_r_i_th}
\end{equation}
\begin{equation}
V_0(r) = V_\mathrm{max} \tanh\left[ \pi \cdot (|R| / R_0) \right]
\label{eq:V0_r}
\end{equation}
We fixed the inclinations ($i$) from $b/a$ ratios
{by modelling the system as an oblate spheroid with an intrinsic axis ratio $q_0$ of 0.2, so that the inclination $i$ satisfies: $$\cos^2 i = \left( (b/a)^2 - q_0^2 \right) / (1 - q_0^2),$$
\citep{hubble26}}. We fixed $R_0$ as $R_\textrm{eff}$ and central positions ($x_0$, $y_0$) derived from the extended components of the galaxies in the photometric analysis (\ref{subsec:galfit_phot}). We used the parameters obtained for b and c galaxies from the exponential disk component because the S\'ersic components are oriented close to face-on with small $R_\textrm{eff}$. The parameters $V_\mathrm{max}$, $V_\mathrm{sys}$, and position angle (P.A.) were fitted. The fitted stellar kinematics parameters are in good agreement with the gaseous kinematics (Table~\ref{tab:Gal_props}). 

The final step in determining the mass-to-light ratio is the estimation of the stellar population properties in each galaxy. As shown in Fig.~\ref{fig:LOSVD_bins}, some spaxels contain spectral contributions from multiple galaxies, which we can now divide using the previously derived kinematic initial approximations. In our analysis, we used the \nb\ extension to perform spectral fitting with two stellar components. The key difference between this method and the one-component approach is that here the spectrum is modeled by the sum of two independent stellar populations, each with its own parametric LOSVD. This approach is particularly useful when there are significant differences in the velocities of the stellar populations and a comparable contribution to the galaxy luminosity, as is the case in systems with counter-rotation phenomenon \citep[e.g.,][]{Katkov2013ApJ...769..105K,Katkov2016MNRAS.461.2068K,Katkov2024ApJ...962...27K}. Our merging system is another great example for applying this technique.

However, this method is highly sensitive to both the signal-to-noise ratio and the accuracy of the initial parameter estimates, making it unsuitable for automatic fit in all spaxels. Hence, we selected several high SNR spaxels (3-4 in each case) that feature contributions from only the a$_1$ component, as well as spaxels with contributions from both a$_1$ and other components. We then fitted these spaxels using the appropriate number of stellar populations. Performing multiple fits allows us to cross-check the results and avoid false $\chi^2$ minima. The initial kinematic estimates were derived from the “circular” rotation model, and these velocities were confirmed. The mean ages and metallicities of the stellar populations are listed in Table~\ref{tab:losvd_phot}. Using these values, we computed the mass-to-light ratios in the SDSS r-band.

\subsection{Stellar masses}
\label{subsec:galfit_phot}

Stellar masses were derived from the luminosities of the galaxies obtained by the Galfit analysis and the mass-to-light ratios. For the galaxies b and c, we combined the integrated magnitudes of their components, while for a$_1$ and a$_2$, we used the integrated magnitude of the S\'ersic component. Assuming a luminosity distance of 428.6 Mpc (for z = 0.0935275), we calculated the luminosities and stellar masses of all the galaxies (Table~\ref{tab:losvd_phot}).


\begin{table*}[h]
\centering
\small
\begin{tabular}{l|ccccccc}
 \hline
 Galaxy & a$_1$ & a$_2$ & a$_{\text{br}}$ & \multicolumn{2}{c}{b}  & \multicolumn{2}{c}{c} \\ 
 Component & S\'ersic & S\'ersic & Exp. disk & S\'ersic & Exp. disk & S\'ersic & Exp. disk \\ \hline 
 R.A. (J2000), 15$^h$58$^m$50$^s$ +  & 0.45$^s$ & 0.34$^s$ & 0.39$^s$ & 0.33$^s$ & 0.22$^s$ & 0.65$^s$ & 0.74$^s$ \\
 
 Dec. (J2000), 27$^\circ$23$'$20$''$ + & 4.6$''$ & 2.6$''$ & 3.3$''$ & 8.0$''$ & 9.0$''$ & 9.3$''$ & 10.7$''$ \\ 
 Offset, arcsec & --- & --- & --- & \multicolumn{2}{c}{$1.74 \pm 0.02$} & \multicolumn{2}{c}{$1.75 \pm 0.02$} \\
 P.A.$_{\textrm{phot}}$ & $(29.9 \pm 0.3)^\circ$ & $(-25 \pm 1)^\circ$ & $(40.7 \pm 0.4)^\circ$ & $(12 \pm 8)^\circ$ & $(64 \pm 4)^\circ$ & $(-14 \pm 4)^\circ$ & $(-32 \pm 1)^\circ$ \\ 
 R$_{\textrm{eff}}$, arcsec & $3.56 \pm 0.08$ & $3.89 \pm 0.92$ & $0.80 \pm 0.01$ & $0.64 \pm 0.01$ & $3.0 \pm 0.03$ & $0.61 \pm 0.01$ & $2.37 \pm 0.04$ \\ 
 b/a & $0.597 \pm 0.002$ & $0.622 \pm 0.007$ & $0.61 \pm 0.01$ & $0.967 \pm 0.005$ & $0.823 \pm 0.008$ & $0.878 \pm 0.008$ & $0.496 \pm 0.008$ \\ 
 m$_{r,\textrm{comp}}$ & $16.11 \pm 0.02$ & $17.36 \pm 0.10$ & $17.52 \pm 0.01$ & $17.99 \pm 0.01$ & $18.01 \pm 0.02$ & $18.99 \pm 0.01$ & $18.94 \pm 0.01$ \\ 
 m$_{r,\textrm{total}}$ & $16.11 \pm 0.02$ & $17.36 \pm 0.10$ &  $17.52 \pm 0.01$ &\multicolumn{2}{c}{$17.25 \pm 0.02$} & \multicolumn{2}{c}{$18.2 \pm 0.02$} \\ \hline
 Age,~Gyr & $1.5$ & $0.4$ & --- & \multicolumn{2}{c}{$2.3$} & \multicolumn{2}{c}{$2.0$} \\
 $[$Fe/H$]$, dex & $-1.1$ & $-0.7$ & --- & \multicolumn{2}{c}{$-0.1$} & \multicolumn{2}{c}{$-0.6$} \\
 $(M/L)$, $(M_\odot/L_\odot)$ & 0.44 & 0.28 & --- & \multicolumn{2}{c}{0.90} & \multicolumn{2}{c}{0.63} \\
 M$_\star$, $10^{9}$\Ms & $21.0 \pm 0.3$ & $4.1 \pm 0.4$ & --- & \multicolumn{2}{c}{$15.0 \pm 0.2$} & \multicolumn{2}{c}{$4.3 \pm 0.1$} \\\hline
\end{tabular}
\caption{Parameters obtained from stellar population fits (see Subsec. \ref{subsec:St_pop_method}). First part of the Table contains parameters from photometric analysis: Right Ascension (R.A.) and Declination (Dec.) were calculated from the center positions of “cores” in the \textsc{galfit} analysis in the CFHT MegaPipe $r$ image; P.A., R$_{\textrm{eff}}$, and b/a are the position angle, effective radius and axis ratio of the extended component, and m$_{r,\textrm{total}}$ is the total magnitude of the galaxy. Second part of the Table shows parameters of stellar populations of galaxies: age and metallicity $[$Fe/H$]$ were obtained from a two-component \nb\ analysis, the mass-to-light $(M/L)$ ratio was calculated based on these stellar population parameters in the $r$ filter, and M$_\star$ is the stellar mass of the components, which was calculated from luminosity and the mass-to-light ratio in the $r$ band.}
\label{tab:losvd_phot}
\end{table*}

\section{Gas Analysis}
\label{sect:gas}
To analyze the gas content within the group, we used the Hybrid-10 Model {\sc logcube} made by the DAP from the MILESHC template for the stellar kinematics and the MaSTAR stellar library. Emission-line properties remain unbinned, while stellar properties are VORONOI binned to a signal-to-noise ratio of approximately 10. We subtracted the fits of emission-lines from the fits of the full spectra to obtain the stellar contribution to the spectra and we subtracted this contribution from the observed spectra to isolate the emission originating from the gas. We applied the correction provided by MaNGA for Galactic extinction, as a function of wavelength. The rest of the analysis described in this section is based on these spectra. 

\begin{figure}[h]
\centering
    \includegraphics[scale =0.57, keepaspectratio]{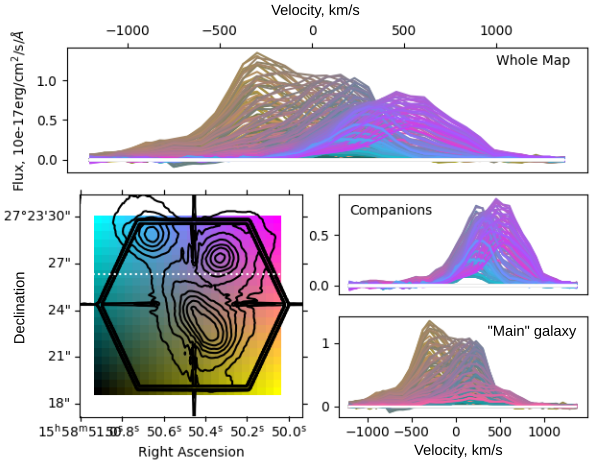}
    \caption{O[{\sc iii}] emission lines per pixel, colour-coded by the position of the pixel (as shown in the left panel). The SDSS g-band image is plotted as black contours on the bottom left panel and the hexagon is the field of view of MaNGA. The spectra of the whole map have been plotted in the top panel. The middle right plot illustrates the top half of the field lying above the dotted white line, emphasizing regions where companion galaxies are present. Conversely, the bottom right plot portrays the bottom half of the field below the white dotted line, highlighting areas where the `main'' galaxy $a_1$ is situated.}
    \label{fig:OIII_emlines}
\end{figure}

Fig.~\ref{fig:OIII_emlines} illustrates the large width of the [OIII] emission-line and the variety of shapes depending on the region. Given the width of emission lines, the H$\alpha$ line is blended with the [NII] doublet, complicating the identification of individual components. While the DAP offers single-Gaussian fits for each emission line, significant residual spectra remain. These residuals call for a fitting of gas emission-lines from a model with several galaxies. Residuals due to the presence of intra-group components, such as tidal tails, are expected to remain. 

\begin{table}
    \centering
    \small
    \begin{tabular}{l|cccc}
    \hline
    & \multicolumn{4}{|c}{Stellar kinematics} \\
    Galaxy & a$_1$ & a$_2$ & b & c \\
    \hline
     R.A., & & & & \\
     15$^h$58$^m$50$^s$ + & \textit{0.45$^s$} & \textit{0.34$^s$} & \textit{0.22$^s$} & \textit{0.74$^s$} \\
     Dec., & & & & \\
     27$^\circ$23$'$20$''$ + & \textit{4.6$''$} & \textit{2.6$''$} & \textit{9.0$''$} & \textit{10.7$''$} \\
     $V_{\textrm{sys}}$& -100 & $120 \pm 5$ & $481 \pm 10$ & $178 \pm 33$   \\
     
     $V_{\textrm{max}}$, km~s$^{-1}$ & 143 $\pm 3$ & 86 $\pm 7$ & 242 $\pm 17$  & 80 $\pm 27$   \\
     $R_0$, arcsec & \textit{3.56} & \textit{3.89} & \textit{3.0} & \textit{2.36} \\
     P.A. & $(23 \pm 1)^\circ$ & $(322 \pm 4)^\circ$ & $(302 \pm 2)^\circ$ & $(344 \pm 24)^\circ$ \\
     
     $i$, deg & \textit{55$^\circ$} & \textit{53$^\circ$} & \textit{35$^\circ$} &  \textit{62$^\circ$} \\
    \hline
    \end{tabular}
    \caption{Galaxy properties derived from kinematic modelling of recovered stellar velocity fields (for stellar kinematic Eq.~\ref{eq:V_r_i_th}, \ref{eq:V0_r}): R.A. and Dec. are coordinates of the velocity field, $V_\mathrm{sys}$ is the systematic velocity relative to the redshift velocity, $V_{0} \approx 28~070$ km~s$^{-1}$, $V_\mathrm{max}$ is the rotational velocity, $R_0$ is the characteristic radius of the rotation curve, P.A. is the position angle of the velocity field, and $i$ is the inclination of the galaxy. Italic font indicates that this parameter was fixed in the analysis to resolve degeneracy. The two last lines show the possibility of more inclined galaxies. }
    \label{tab:Gal_props}
\end{table}

\subsection{Emission Line Fitting from modelled galaxies}

\subsubsection{Kinematics}

We model the gas kinematics through mock observations of four simulated galactic discs. Each disc is generated with particles having positions and velocities drawn from chosen distributions. We use a Miyamoto-Nagai density profile:
\begin{equation}\label{eq:mn}
\rho_d(R,z) = \left( \dfrac{h^2 M}{4 \pi} \right) \dfrac{a R^2 + (a + 3\sqrt{z^2 + h^2}) (a + \sqrt{z^2 + h^2})^2}
    {\left[ R^2 + (a + \sqrt{z^2 + h^2})^2\right]^{\frac{5}{2}} (z^2 + h^2)^{\frac{3}{2}}},
\end{equation}
where $M$ is the total mass of the disc, $a$ is a radial scale length, and $h$ is a vertical scale length. $a$ is set to 1~kpc for all discs, and $h$ to 0.2~kpc. Discs are cut at four times their radial scale length ($a$). The impact of the exact profile on the modelling is diminished by the following fit in each spaxel, explained below. The rotation curve follows Eq.~\ref{eq:V0_r}, as for the stellar disc. 
The disc is placed at the location of the core component found by the photometric fit\footnote{For galaxies $b$ and $c$, this differs from the stellar fit for which the discs are centred on the centres of the "extended" components. However, for galaxy $b$, emission-line fluxes are the largest in the MaNGA pixel containing the centre of the core component, making it logical to place the centre of the simulated disc there, and we similarly use the centre of the core component for galaxy $c$, which is too close to the edges of the field-of-view to check fluxes near the centre of the extended component.} Inclination, position-angle, line-of-sight (los) systemic velocity, $V_{\rm max}$ (plateau velocity of the rotation curve), $R_0$ (transition radius of the rotation curve) and velocity dispersion (modelled with a single value for uniform and equal radial $\sigma_R$ and azimuthal $\sigma_{\phi}$ velocity dispersions and a uniform vertical velocity dispersion $\sigma_z = 1/2 \sigma_R$) are all free parameters. Because of the high degeneracy, especially due to the presence of four galaxies in a region of only a few PSF wide (the FWHM of the MaNGA PSF in the $g$ and $r$-band is 2.6 arcseconds), we fit these parameters of all discs by eye, guided by the values obtained for the stellar components. 

We realise one mock cube per galaxy, by binning on-sky positions and line-of-sight velocities of the simulated disc particles. Los velocities are initially binned on channels of width 50~km/s and, for a given emission-line, the simulated spectra are interpolated on the velocities obtained using the following conversion between wavelength $\lambda$ and los velocity $v$ :
\begin{equation}
     v = c\,(z - z_{\mathrm{source}})/(1 + z_{\mathrm{source}}), \,  \text{with} \, z = \lambda/\lambda_{\mathrm{0}} - 1,
    \label{Eq_v_z}
\end{equation}
where $z_{\rm source}$ is the redshift and $\lambda_0$ is the rest-frame wavelength  in vaccum of the considered emission-line, collected from the NIST Atomic Spectra Database\footnote[1]{https://physics.nist.gov/PhysRefData/ASD/lines\_form.html}. At each simulated spaxel, we perform a convolution on velocities with a 1D Gaussian kernel with the dispersion of the MaNGA line-spread-function for the spaxel at the redshifted wavelength of the considered emission-line (this dispersion is given in the MaNGA model cube). The planes corresponding to each simulated velocity channels are convolved by a circular Gaussian kernel of FWHM 2.6 arcseconds, accounting for the MaNGA PSF in the $g$ and $r$-bands. 

To fit an emission-line from the four simulated mock cubes (with one mock cube per galaxy), we consider a varying contribution of each disc in each spaxel. With $s_i$ the mock spectrum of disc$_i$ ($i=1..4$), we fit four multiplying factors $f_i$, one for each spectrum, so that the sum of the resulting spectra fits best the observed spectrum, i.e. so that $\sum_i f_i s_i$ fits the observed spectrum. As the fitting is performed after convolution of the mock cube by the MaNGA PSF, the fitted factors must not vary strongly on the area of the PSF for this approach to be physically valid (otherwise, the contribution of the gas of a given disc at a given location would be inconsistent from one spaxel to its neighbours on which it is smeared by the spatial convolution). 

We fit the following emission-lines: H$\beta$, [OIII]5008, H$\alpha$, the [NII] doublet and the [SII]6718,6733 doublet. 

We note that, without any clear inclination of disc components derived from the optical images, the large size of the MaNGA PSF compared to the galaxies makes the determination of the plateau rotation velocity and inclination of each disc completely degenerate in our modelling, because the iso-los-velocity curves of a disc are almost straight lines parallel to its minor axis. It is thus only possible to determine $V_{\rm max} \sin i$. We compute the fits with couples of $V_{\rm max}$ and inclination which are reasonable (we especially avoid unrealistically high rotation velocities) and use the inclination to draw the disc on the different figures. 

Low velocity gas (with a los velocity $<500$~km/s) cannot be fitted by our model, as visible on Fig.~\ref{fig:emlines-fit}. This gas may be modelled by a broad kinematic component, of velocity dispersion 500~km/s, spatially and kinematically centred on $a_2$. This additional component is name $a_{\rm 2 \, bis}$. 

\begin{figure*}
    \centering
    \includegraphics[width = \linewidth]{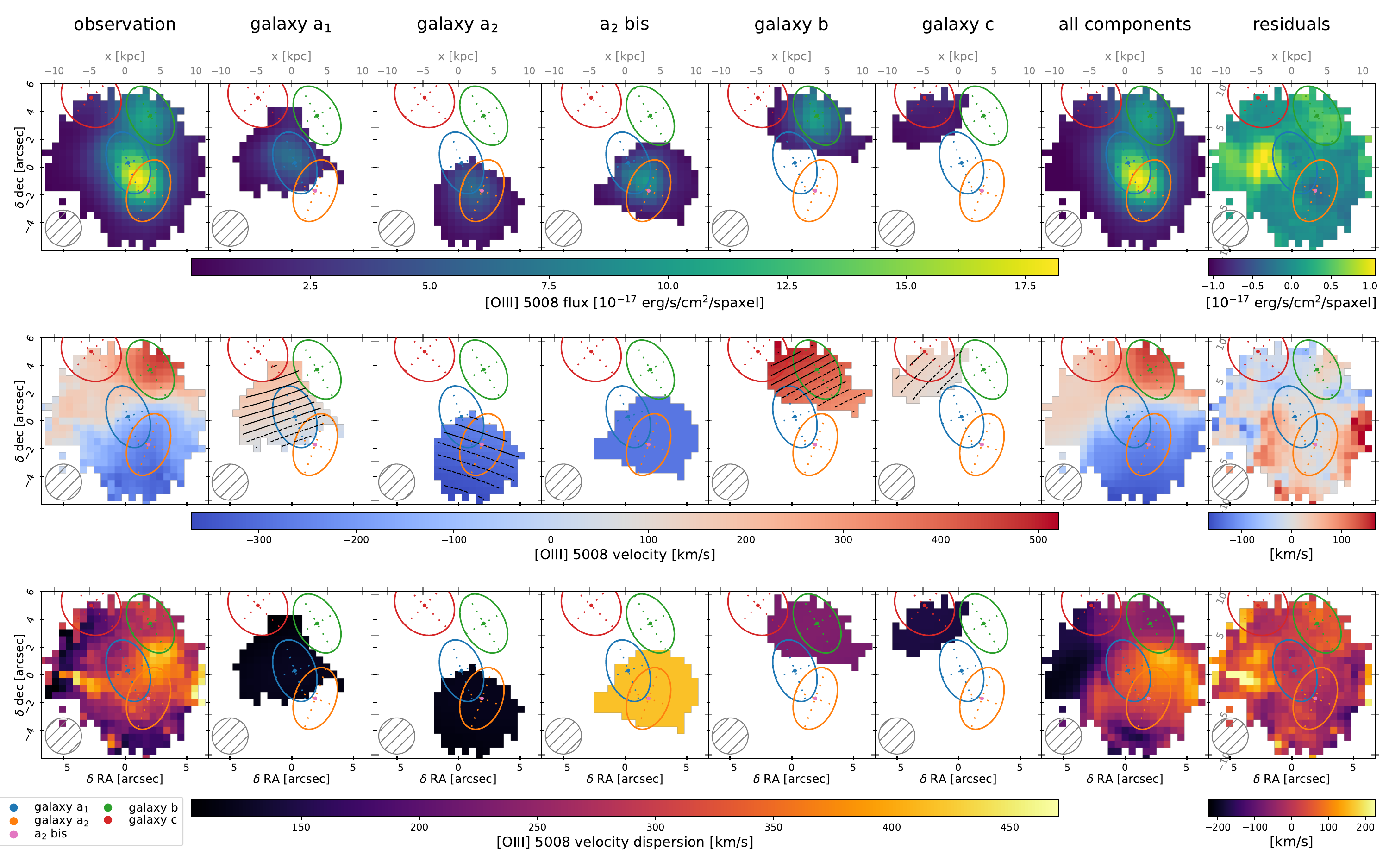}
    \caption{Observations, models and residuals of flux, velocity and velocity dispersion from the [OIII] 5008 Angstrom line. Models are shown both for each component (column 2 to 6) and for the 5 components together (column 7). Residuals (column 8) are the observation (column 1) minus the model (column 7). On the velocity maps of individual galaxies, iso-los velocity curves are represented in black dashed (resp. solid) lines for values below (resp. above) the systemic velocity of each galaxy.}
    \label{fig:moms-res-oiii-fit}
\end{figure*}

\begin{figure*}
    \centering
    \includegraphics[width = \linewidth]{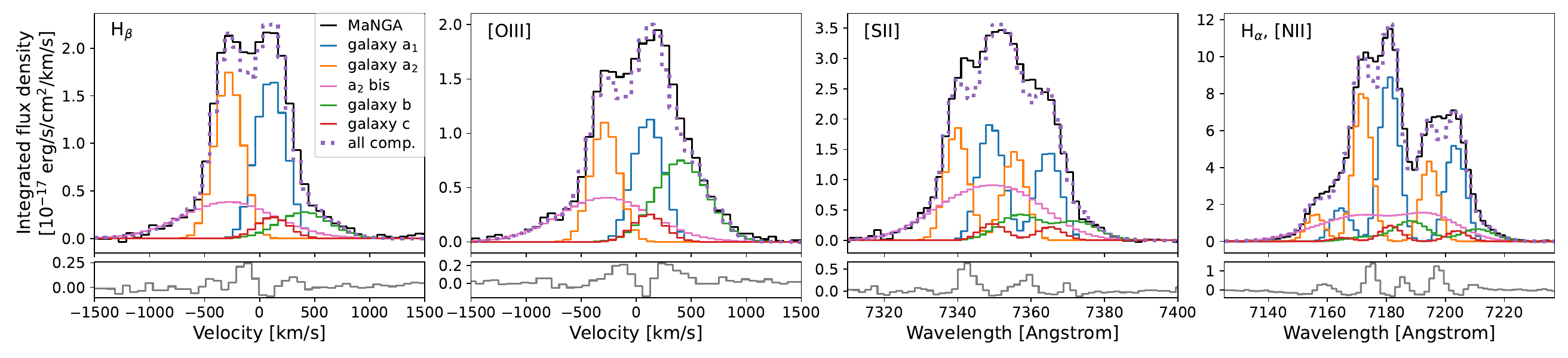}
    \caption{Observations and models of emission-lines integrated on the whole system, with residuals below each panel.}
    \label{fig:emlines-fit}
\end{figure*}

Figs~\ref{fig:spectra-oiii-fit} shows observed and modelled spectra for one of the fitted emission-lines, [OIII]5008, stacked 16 by 16 at the corresponding on-sky locations, with the MegaPipe image as background. Figs~\ref{fig:moms-res-oiii-fit}  shows the moments for the [OIII]5008 line. Fig~\ref{fig:emlines-fit} shows the emission-lines stacked on the whole system in the observation and the models. Parameters used for the models are given in Table \ref{tab:Gas_Gal_props}. While systemic velocities of galaxies $b$ and $c$ are close to the ones found in the stellar analysis, we obtain very different results for galaxies $a_1$ and $a_2$. Emission-lines peak near $100$~km/s at the location of the centre of $a_1$, which makes us attribute this systemic velocity to $a_1$. A neighbouring galaxy with a lower systemic velocity is needed to fit the emission-lines, hence the systemic velocity $-250$~km/s attributed to $a_2$. The differences in the results found by the stellar and gas analyses for these two galaxies reflect the high degeneracy of the system and the difficulties created by the low spatial resolution of the observation compared to the sizes of the galaxies.  

The velocity dispersions which are required to fit the spectra are especially large : 80~km/s for galaxies~$a_1$ and 100~km/s for galaxy $a_2$, and as much as 250~km/s for galaxy~$b$ and 180~km/s for galaxy~$c$. While the values for galaxies $a_1$
 and $a_2$ may be attributed to kinematically hot gas in perturbed discs, the larger values for $b$ and $c$ likely reflect either an AGN broadline origin for $b$, or that the emission-lines are due to several galaxies. As previously mentioned, galaxy $c$ is indeed close to a galaxy lying just outside the MaNGA field-of-view, and galaxy $c$ might consist of more than one galaxy, as may be best guessed from the Subaru images of Fig.~\ref{fig:subaru}.

\begin{table}
    \centering
    \begin{tabular}{l|ccccc}
    \hline
    & \multicolumn{5}{|c}{Gas kinematics} \\
    Component & a$_1$ & a$_2$ & a$_{\rm 2, bis}$ & b & c  \\
    \hline
     R.A., 15$^h$58$^m$50$^s$ + & 0.45$^s$ & 0.34$^s$ & 0.34$^s$ & 0.33$^s$ & 0.65$^s$ \\
     Dec., 27$^\circ$23$'$20$''$ + & 4.6$''$ & 2.6$''$ & 2.6$''$ & 8.0$''$ & 9.3$''$ \\
     $V_{\textrm{sys}}$, km~s$^{-1}$ & 100 & -280 & -280 & 425 & 150 \\
     $V_{\textrm{max}}$, km~s$^{-1}$ & 200 & 140 &  & 200 & 200  \\
     $R_0$, kpc & 3.14 & 3.14  & & 3.14 & 3.14  \\
     $\sigma_{R}$, km~s$^{-1}$ & 80 & 100 & 500 & 250 & 180 \\
     P.A., deg & $20^\circ$ & $340^\circ$ & $340^\circ$ & $30^\circ$ & $45^\circ$ \\
     $i$, deg & 50$^\circ$ &50$^\circ$ & 50$^\circ$ & 50$^\circ$ &  30$^\circ$\\
     
     $V_{\textrm{max}} \sin i$, km~s$^{-1}$ & 153 & 107 &  & 153  & 100  \\
    \hline
    \end{tabular}
    \caption{Gas modelling. Here $V_{\textrm{sys}}$ is defined with Eq.~\ref{Eq_v_z}.}
    \label{tab:Gas_Gal_props}
\end{table}

\subsubsection{Star formation rate and BPT diagrams from the modelling}

We compute an H$\alpha$ based SFR using \citet{1998ARA&A..36..189K}:
\begin{equation}
{\rm SFR(H\alpha)}  =  7.9 \, 10 ^{-42} \, \frac{L_{\rm corr}(\mathrm{H\alpha)}}{\rm erg / s} \mathrm{M_{\odot}/yr},
\end{equation}
which assumes a Salpeter initial mass function. $L_{\rm corr}(H\alpha)$ is the luminosity of the H$\alpha$ emission-line corrected from dust attenuation. $L_{\rm corr}(\rm H\alpha)$ is obtained by correcting the observed luminosity $L_{\rm obs}\rm (H\alpha)$ by:
\begin{equation}
    L_{\rm corr}(\mathrm{H\alpha}) = L_{\rm obs}(\mathrm{H\alpha)}\, 10^{0.4 \, k(\lambda_{H\alpha}) \, E(B-V)},
    \label{eq-dustcorr}   
\end{equation}
where $k(\lambda_{\rm H\alpha}) = R_V A_{\lambda_{\rm H\alpha}} / A_V $ is the value of the total attenuation curve at the H${\alpha}$ rest-frame emission wavelength, with $R_V$ the ratio of total to selective extinction in the $V$ band, $A_{\lambda_{\rm H\alpha}}$ and $A_V$ the attenuations at the H$_{\alpha}$ wavelength and in the $V$ band (respectively), and $E(B-V) = A_B$ - $A_V$ the colour excess. Using Eq.~\ref{eq-dustcorr} applied to both H$\alpha$ and H$\beta$, the colour excess is computed by:
\begin{equation}
    E(B-V) = \dfrac{2.5}{k(\lambda_{\rm H\beta}) - k(\lambda_{\rm H\alpha)})} \, \mathrm{log}_{10} \left[ \frac{\rm (H\alpha / H\beta)_{obs}}{2.86}\right],
    \label{eq-ebmv}
\end{equation}
where ${\rm (H\alpha / H\beta)_{obs}}$ is the Balmer decrement, the ratio of observed H$\alpha$ and H$\beta$ luminosities, and an intrinsic Balmer decrement H$\alpha$/H$\beta$ of 2.86 is assumed, which corresponds to a temperature of ${\rm T=10^4\,K}$ and an electron density of $n_e=10^2{\rm cm^{-3}}$, for a case B recombination \citep{Osterbrock2006}. We use the total attenuation curve parametrised by \citet{Odonnel94}, with $R_V=3.1$. 

\begin{figure*}
    \centering
    \begin{tabular}{cc}
       
    \includegraphics[width = 0.47 \linewidth]{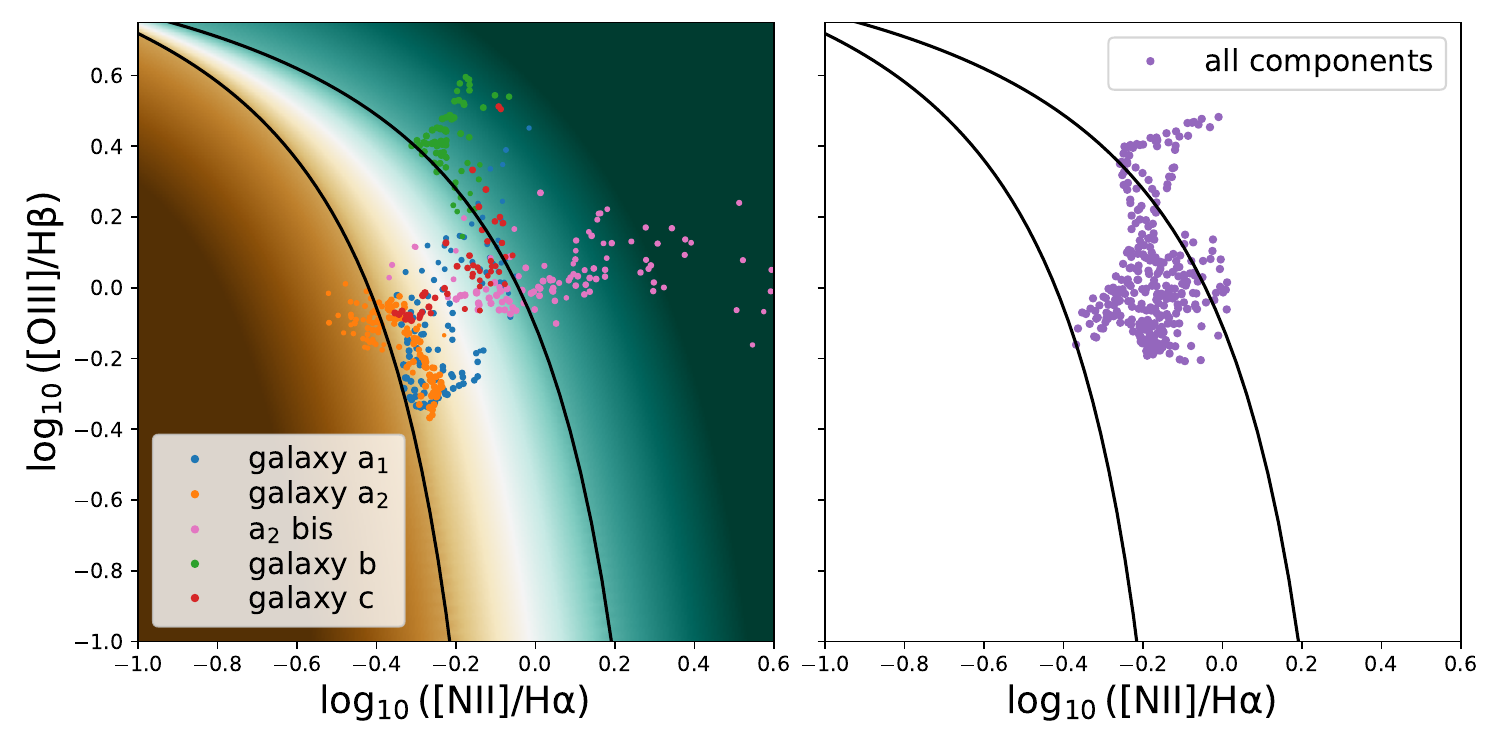}  & 
    \includegraphics[width = 0.47 \linewidth]{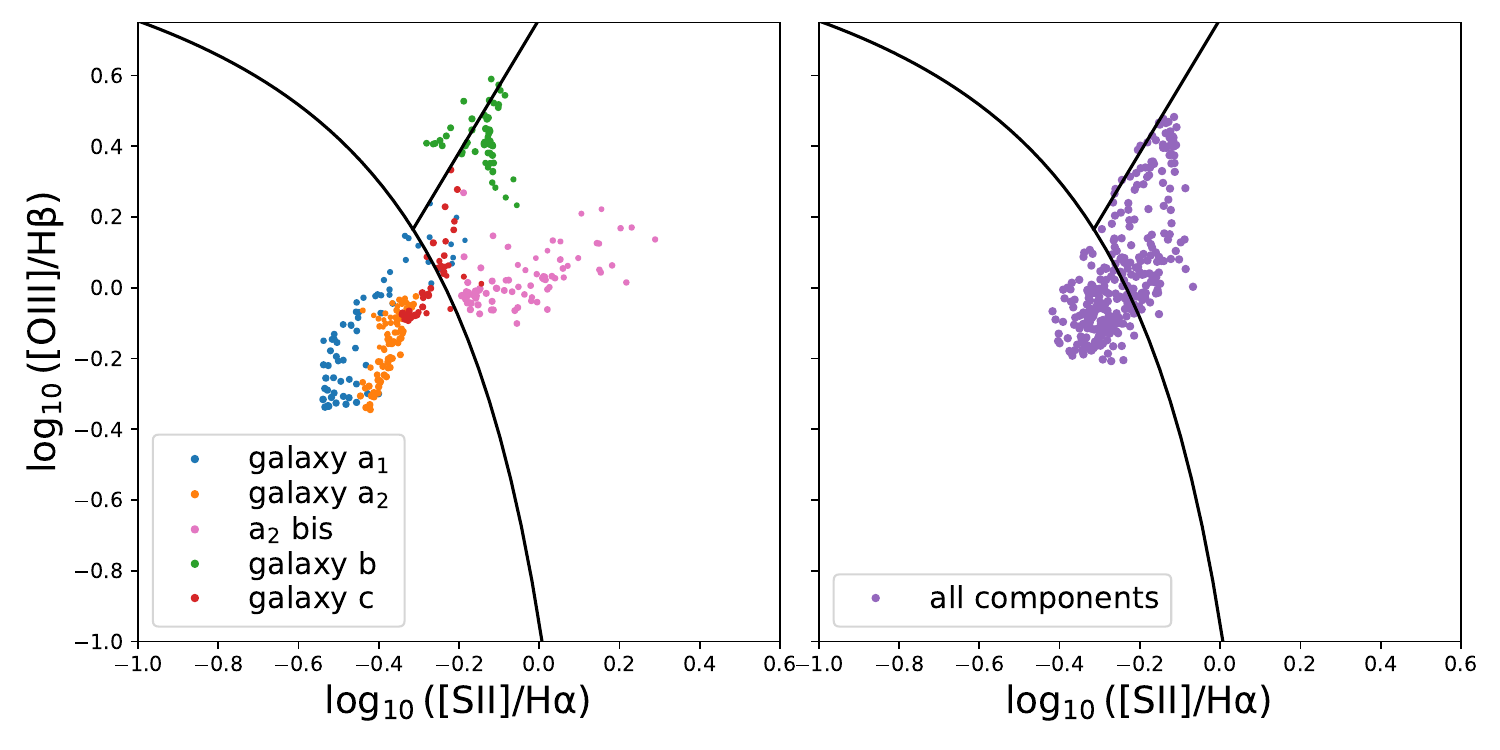} 
    
    \end{tabular}
    \includegraphics[width = \linewidth]{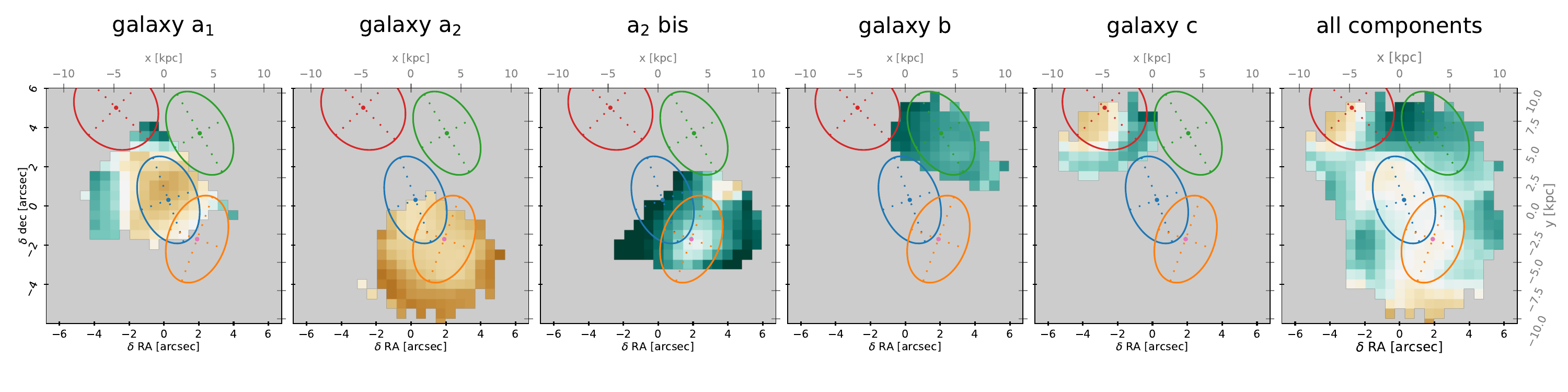}
    \caption{BPT diagrams. Top row: From left ro right: [NII] BPT for the 5 individual components, [NII] BPT for all the components together, [SII] BPT for the 5 individual components, [SII] BPT for all the components together. Bottom row: maps with each pixel colout-coded according to the location on the BPT [NII] diagram coloured on the top left panel. Ellipses (or single dot for $a_{\rm 2, bis}$ show the locations of the different components, with the same colour code as in Fig.~\ref{fig:moms-res-oiii-fit}.} 
    \label{fig:bpt-fit}
\end{figure*}

By computing the H$\alpha$ and H$\beta$ fluxes per spaxel and per modelled galaxy and applying the correction for attenuation of Eq.\ref{eq-dustcorr}, we can obtain a SFR map per galaxy. The total SFR per galaxy is shown in Table~\ref{tab:SFR}. Using the [SII] doublet and the [OIII] emission-line, it is possible to correct for the contribution of AGN to the H$\alpha$ flux, following \citet{Jin2021}. The corrected SFRs are shown in Table~\ref{tab:SFR}. Last, Fig~\ref{fig:bpt-fit} show the BPT diagrams obtained both for the whole system and by separating the emissions of individual galaxies.

\begin{table}
    \centering
    \begin{tabular}{l|cccccc}
    \hline
    Galaxy & a$_1$ & a$_2$ & a$_{\rm 2, \, bis}$ & b & c & all \\
    \hline
     SFR$_{raw}$, M$_{\odot}$/yr  & 23 & 17 & 8 & 3 & 1 & 52\\
     \hline
     SFR$_{corr}$, M$_{\odot}$/yr  & 19 & 13 & 4 & 1 & 1 & 45\\
    \hline
    \end{tabular}
    \caption{SFR per galaxy from our modelling, with an additional broad component, a$_{\rm 2, \, bis}$. The first (resp. second) line provides the SFR$_{raw}$ (resp. SFR$_{corr}$) without (resp. with) AGN correction.}
    \label{tab:SFR}
\end{table}

\section{Discussion}
\label{sect:disc}
\begin{figure}
    \centering
    \fbox{\begin{adjustbox}{width=0.46\linewidth,keepaspectratio}
\begin{tikzpicture}%
[x=-0.5cm,y=0.5cm,z=-0.3cm,>=stealth,scale=0.5]
\coordinate (aa) at (0.,0.,0);
\coordinate (ab) at (-1.47,-2,0);
\def\abx{-1.47};
\def\aby{-2.};
\coordinate (ba) at (-1.61,3.4,0);
\def\bax{-1.61};
\def\bay{3.4};
\coordinate (bb) at (-3.07,4.4,0);
\def\bbx{-3.07};
\def\bby{4.4};
\coordinate (ca) at (2.66,4.7,0);
\def\cax{2.66};
\def\cay{4.7};
\coordinate (cb) at (3.86,6.1,0);
\def\cbx{3.86};
\def\cby{6.1};

\def\ppaaa{20.+90.};
\def\ppaab{340.0+90.};
\def\ppaba{30.+90.};
\def\ppaca{45.+90.};
\def\boaaa{0.66};
\def\boaab{0.66};
\def\boaba{0.66};
\def\boaca{0.87};
\def\iaa{acos(sqrt((\boaaa*\boaaa-0.2*0.2)/(.-0.2*0.2)))};
\def\iab{acos(sqrt((\boaab*\boaab-0.2*0.2)/(.-0.2*0.2)))};
\def\iba{acos(sqrt((\boaba*\boaba-0.2*0.2)/(.-0.2*0.2)))};
\def\ica{acos(sqrt((\boaca*\boaca-0.2*0.2)/(.-0.2*0.2)))};
\def\reaa{3.14/1.4}; 
\def\reab{3.14/1.4};
\def\reba{3.14/1.4};
\def\reca{3.14/1.4};

\def\baa{\boaaa*\reaa};
\def\bab{\boaab*\reab};
\def\bba{\boaba*\reba};
\def\bbb{\boabb*\rebb};
\def\bca{\boaca*\reca};
\def\bcb{\boacb*\recb};

\def\mhexax{3.993*0.5-11.998/2.-3.993*0.5};
\def\mhexay{16.001*0.5-10.657/2.-5.346*0.5};
\def\mhexbx{8.968*0.5-11.998/2.-3.993*0.5};
\def\mhexby{26.659*0.5-10.657/2.-5.346*0.5};
\def\mhexcx{23.014*0.5-11.998/2-3.993*0.5};
\def\mhexcy{26.659*0.5-10.657/2-5.346*0.5};
\def\mhexdx{27.989*0.5-11.998/2-3.993*0.5};
\def\mhexdy{16.001*0.5-10.657/2-5.346*0.5};
\def\mhexex{23.014*0.5-11.998/2-3.993*0.5};
\def\mhexey{5.346*0.5-10.657/2-5.346*0.5};
\def\mhexfx{8.968*0.5-11.998/2-3.993*0.5};
\def\mhexfy{5.346*0.5-10.657/2-5.346*0.5};
\draw[->] (canvas cs: x=4.5cm) -- (canvas cs: x=-4.25cm) node[above] {$RA$};
\draw[->] (canvas cs: y=-4.25cm) -- (canvas cs: y=4.25cm) node[right] {$DEC$};
\draw[purple,line width=0.5mm] (\mhexax,\mhexay) -- (\mhexbx,\mhexby) -- (\mhexcx,\mhexcy) -- (\mhexdx,\mhexdy) -- (\mhexex,\mhexey) -- (\mhexfx,\mhexfy) -- (\mhexax,\mhexay) ; 
\foreach \coo in {-8,-7,...,5,6,7,8}
{
  \draw (\coo,-1.5pt) -- (\coo,1.5pt);
  \draw (-1.5pt,\coo) -- (1.5pt,\coo);
}

\foreach \coo in {-5,5}
{
  \draw[thick] (\coo,-3pt) -- (\coo,3pt) node[below=3pt] {\coo};
  \draw[thick] (-3pt,\coo) -- (3pt,\coo) node[left=3pt] {\coo};
}
\draw[rotate=\ppaaa] (aa) ellipse ({\reaa} and {\baa});
\node[fill, circle, scale=0.3, black,label={[xshift=-0.8cm, yshift=-0.2cm]right:$a_1$}] at (aa) {};
\node[fill, circle, scale=0.3, black, label={[xshift=.4cm, yshift=-0.4cm]above:$a_2$}] at (ab) {};
\draw [rotate around={\ppaab:(ab)},x radius={\reab},y radius={\bab}] (ab) ellipse;
\node[fill,circle,scale=0.3,black,label={[xshift=.0cm, yshift=0.1cm]above:$b$}] at (ba) {};
\draw[rotate around={\ppaba:(ba)}] (ba)  ellipse ({\reba} and {\bba});
\node[fill,circle,scale=0.3,black,label={[xshift=.0cm, yshift=0.1cm]above:$c$}] at (ca) {};
\draw[rotate around={\ppaca:(ca)}] (ca)  ellipse ({\reca} and {\bca});
\end{tikzpicture}
    \end{adjustbox}}
    \fbox{\begin{adjustbox}{width=0.46\linewidth,keepaspectratio}
\begin{tikzpicture}%
[x=-0.5cm,y=0.5cm,z=-0.3cm,>=stealth,scale=0.5]
\coordinate (aa) at (0.,0.,0);
\coordinate (ab) at (-1.47,-2,0);
\def\abx{-1.47};
\def\aby{-2.};
\coordinate (ba) at (-1.61,3.4,0);
\def\bax{-1.61};
\def\bay{3.4};
\coordinate (bb) at (-3.07,4.4,0);
\def\bbx{-3.07};
\def\bby{4.4};
\coordinate (ca) at (2.66,4.7,0);
\def\cax{2.66};
\def\cay{4.7};
\coordinate (cb) at (3.86,6.1,0);
\def\cbx{3.86};
\def\cby{6.1};

\def\ppaaa{29.9+90.};
\def\ppaab{-25.0+90.};
\def\ppaba{12.+90.};
\def\ppabb{64.+90.};
\def\ppaca{-14.+90.};
\def\ppacb{-32.+90.};

\def\boaaa{0.597};
\def\boaab{0.622};
\def\boaba{0.967};
\def\boabb{0.823};
\def\boaca{0.878};
\def\boacb{0.496};

\def\iaa{acos(sqrt((\boaaa*\boaaa-0.2*0.2)/(.-0.2*0.2)))};
\def\iab{acos(sqrt((\boaab*\boaab-0.2*0.2)/(.-0.2*0.2)))};
\def\iba{acos(sqrt((\boaba*\boaba-0.2*0.2)/(.-0.2*0.2)))};
\def\ibb{acos(sqrt((\boabb*\boabb-0.2*0.2)/(.-0.2*0.2)))};
\def\ica{acos(sqrt((\boaca*\boaca-0.2*0.2)/(.-0.2*0.2)))};
\def\icb{acos(sqrt((\boacb*\boacb-0.2*0.2)/(.-0.2*0.2)))};

\def\reaa{3.56}; 
\def\reab{3.89};
\def\reba{0.64};
\def\rebb{3.0};
\def\reca{0.61};
\def\recb{2.37};

\def\baa{\boaaa*\reaa};
\def\bab{\boaab*\reab};
\def\bba{\boaba*\reba};
\def\bbb{\boabb*\rebb};
\def\bca{\boaca*\reca};
\def\bcb{\boacb*\recb};

\def\mhexax{3.993*0.5-11.998/2.-3.993*0.5};
\def\mhexay{16.001*0.5-10.657/2.-5.346*0.5};
\def\mhexbx{8.968*0.5-11.998/2.-3.993*0.5};
\def\mhexby{26.659*0.5-10.657/2.-5.346*0.5};
\def\mhexcx{23.014*0.5-11.998/2-3.993*0.5};
\def\mhexcy{26.659*0.5-10.657/2-5.346*0.5};
\def\mhexdx{27.989*0.5-11.998/2-3.993*0.5};
\def\mhexdy{16.001*0.5-10.657/2-5.346*0.5};
\def\mhexex{23.014*0.5-11.998/2-3.993*0.5};
\def\mhexey{5.346*0.5-10.657/2-5.346*0.5};
\def\mhexfx{8.968*0.5-11.998/2-3.993*0.5};
\def\mhexfy{5.346*0.5-10.657/2-5.346*0.5};

\draw[->] (canvas cs: x=4.5cm) -- (canvas cs: x=-4.25cm) node[above] {$RA$};
\draw[->] (canvas cs: y=-4.25cm) -- (canvas cs: y=4.25cm) node[right] {$DEC$};

\draw[purple,line width=0.5mm] (\mhexax,\mhexay) -- (\mhexbx,\mhexby) -- (\mhexcx,\mhexcy) -- (\mhexdx,\mhexdy) -- (\mhexex,\mhexey) -- (\mhexfx,\mhexfy) -- (\mhexax,\mhexay) ; 
\foreach \coo in {-8,-7,...,5,6,7,8}
{
  \draw (\coo,-1.5pt) -- (\coo,1.5pt);
  \draw (-1.5pt,\coo) -- (1.5pt,\coo);
}
%
\foreach \coo in {-5,5}
{
  \draw[thick] (\coo,-3pt) -- (\coo,3pt) node[below=3pt] {\coo};
  \draw[thick] (-3pt,\coo) -- (3pt,\coo) node[left=3pt] {\coo};
}
\draw[rotate=\ppaaa] (aa) ellipse ({\reaa} and {\baa});
\node[draw,cross out, minimum size=3pt, inner sep=0pt, black,label={[xshift=-0.8cm, yshift=-0.2cm]right:$a_1$}] at (aa) {};
\node[draw, cross out, minimum size=3pt, inner sep=0pt, label={[xshift=.4cm, yshift=-0.4cm]above:$a_2$}] at (ab) {};
\draw [rotate around={\ppaab:(ab)},x radius={\reab},y radius={\bab}] (ab) ellipse;
\node[draw, cross out, minimum size=3pt, inner sep=0pt,black,label={[xshift=.0cm, yshift=0.1cm]above:$b$}] at (bb) {};
\node[fill,circle,scale=0.3,black] at (ba) {};
\draw[rotate around={\ppaba:(ba)}] (ba)  ellipse ({\reba} and {\bba});
\draw[rotate around={\ppabb:(bb)},dashed] (bb)  ellipse ({\rebb} and {\bbb});
\node[fill,circle,scale=0.3,black] at (ca) {};
\node[draw, cross out, minimum size=3pt, inner sep=0pt,black,label={[xshift=.0cm, yshift=0.1cm]above:$c$}] at (cb) {};
\draw[rotate around={\ppaca:(ca)}] (ca)  ellipse ({\reca} and {\bca});
\draw[rotate around={\ppacb:(cb)},dashed] (cb) ellipse ({\recb} and {\bcb});
\end{tikzpicture}
    \end{adjustbox}}
    \caption{Geometrical configuration. Left: The ellipses correspond to the stellar component of each galaxy, namely its PA and inclination. The four black bullets correspond to the position of the four galaxies on the sky.   Right: Same for the gas component. The dashed lines correspond to the exponential profiles. The bullets and crosses correspond to the conventions in Fig. \ref{fig:galfit_megapipe}.}
    \label{fig:geom1}
\end{figure}
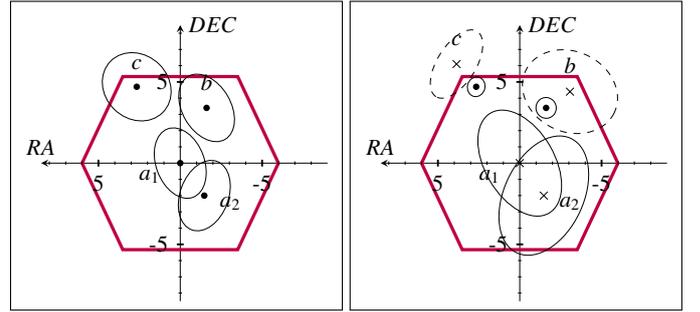

Figure  \ref{fig:geom1} displays the geometrical configuration of the stellar (left) and  gas (right) modelling. We identify four rotating discs that are not yet completely disrupted in the projected field of view of MaNGA (35\,kpc in diameter).
However, we observe a velocity difference of $+200\,$km\,s$^{-1}$ and $-400$\,km\,s$^{-1}$, and we discuss the sensitivity of each galactic disc to the ram-pressure,  stripping that could impact these discs.
The maximum difference of the systemic velocity is $\Delta v_{gas} = 600$\,km/s for the gas and $\Delta v_{star} = 505$\,km/s for the stars, and we study possible tidal effects in Sect. \ref{ssect:tidal}.

\subsection{Ram-pressure stripping}
We first estimate the gravitational restoring force per unit mass $P_{d}$ exerted on the gas content due to the gravitational potential of the galaxy\,:
\begin{equation}
    P_{d} = 2 \pi G\, f_{\mathrm{gas}}\, \Sigma_0^2\, \exp\big(-2 r / r_d\big)
\end{equation}
where $f_{gas}$ is the gas fraction with respect to the total disc mass $M_d$, $\Sigma_0 = M_d / (2\pi r_d^2)$ is the central surface density of the disc, 
$r_d$ is the scale length of the exponential profile, and $r$ the distance from the galactic centre. We assume that the gas fraction is in the range 5-20$\%$.

Second, given the mass of the cluster A2142 
and the projected distance of the galaxy group with respect to the cluster centre ($R_{gg}=1.3$\,Mpc), we estimate the cluster density $\rho$ at the position of the galaxy group as follows. 
We consider a NFW density profile \citep{navarro_universal_1997}, namely:
\begin{equation}
    \rho_{\mathrm{NFW}}(r) = \rho_0 \, (r/r_0)^{-1} \left[ 1 + (r/r_0) \right]^{-2}, 
\end{equation}
with $r_0=R_{200}/c$. {We followed standard conventions: $M_{200} = (4\pi/3)\, \Delta_{200}\, \rho_c\, R_{200}^3$,
$\Delta_{200} = 200$ and $\rho_c = 2.78 \times 10^{11} h^2 M_\odot$\,Mpc$^{-3}$.} 
As summarised in Table \ref{tab:A2142_properties}, we consider $M_{200}=1.25 \times 10^{15} M_\odot$, $R_{200}=2.16$\,Mpc, $r_0=0.54$\,Mpc and $c=4$ \citep[]{2014Munari}. We then calculate: 
\begin{equation}
\rho_0 = \Delta_{200} \, \rho_c \, c^3 \times 3^{-1} \times \left[ \ln(1 + c) - c(1 + c)^{-1} \right]^{-1} 
\end{equation}
and estimate\,: $\rho_0\simeq 7.18 \times 10^{14} \, M_\odot\,\mathrm{Mpc}^{-3}$.
At $r=R_{gg}$, the projected distance of the galaxy group with respect to the cluster centre, it is reasonable to assume that the hot gas profile traces the dark matter NFW profile \citep[e.g.][]{2001MNRAS.327.1353K,Rasia2004}, and therefore 
$\rho_{\mathrm{gas}}(R_{gg}) \sim f_b \, \rho_{\mathrm{NFW}}(R_{gg}) \simeq 3.29 \times 10^{12}\, M_\odot\,\mathrm{Mpc}^{-3}$,
where $f_b = 0.15$ \citep{2016A&A...594A..13P,2016Tchernin} is the (universal) baryonic fraction.
Note that if the group is located at 1.5\,Mpc \citep[e.g.][]{2016Tchernin}, this density would be about 30$\%$ smaller.

As discussed by \citet{2014Eckert}, this galaxy group exhibits a spectacular 700-kpc long X-ray tail, that has last more than 600\,Myr. We can hence estimate the velocity of group in the plane of the sky ($V_{gg}^{sky}<1140$\,km\,s$^{-1}$). In parallel, the radial (or peculiar) velocity can be estimated with the redshift as in Eq. \ref{Eq_v_z} 
with $z_{source}=z_{A2142} = 0.0894$.
We hence estimate for the group redshift $z=0.0934$, a relative velocity with respect to the cluster of 1126\,km s$^{-1}$, and derive in the frame of the A2142 cluster the following (stellar) radial velocities: ${V}_{gal}^{rad}=1026, 1246, 1607$ and $1320 $\,km\,s$^{-1}$ for the a$_1$, a$_2$, b and c galaxies.
\begin{figure}[h]
    \centering
    \includegraphics[width=0.48\textwidth,scale = 1]{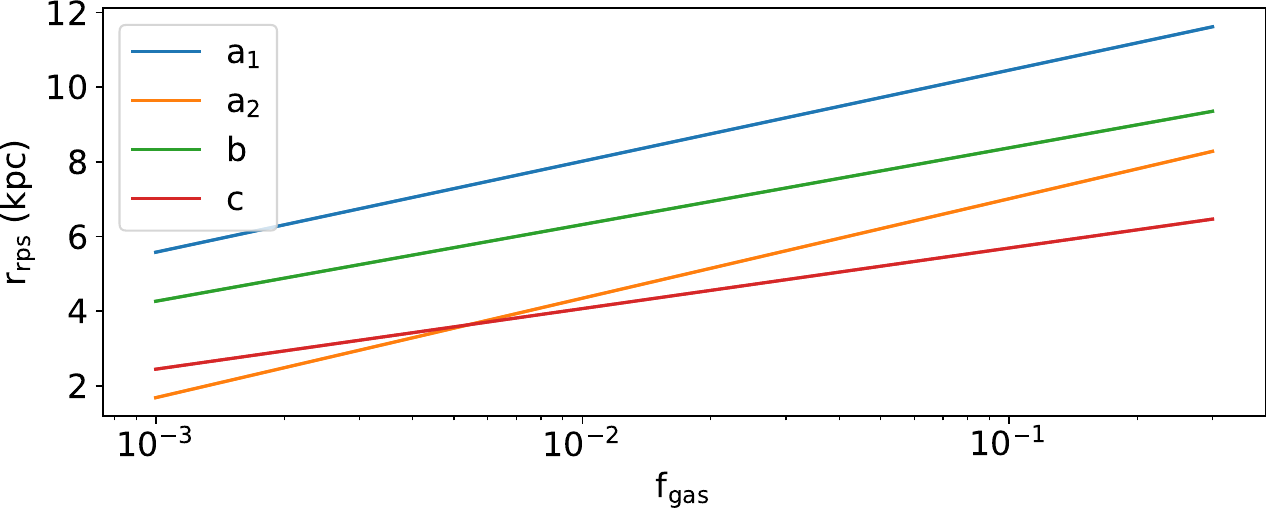}
    \caption{Ram pressure stripping radius as a function of the gas fraction of the galaxy discs. The coloured lines display the values for the disc modelling based on the stellar-derived properties.
    }
    \label{fig:RPS}
\end{figure}
 We thus expect the following total velocities for each member of galaxy group (falling inside the MaNGA field of view): $V_{{a_1}}^{tot}<1530$\,km\,s$^{-1}$, $V_{{a_2}}^{tot}<1690$\,km\,s$^{-1}$, $V_{b}^{tot}<1970$\,km\,s$^{-1}$ and $V_{c}^{tot}<1740$\,km\,s$^{-1}$.


Following \citet{1972ApJ...176....1G}, we compare the ram pressure $\rho_{gas} V_{gal}^2$ to the force per unit mass $P_{disc}$ exerted on the gas of disc-galaxies. We hence estimate the radius $r_{rps}$ at which all the cold gas of the galaxy disc is stripped, with the semi-analytic approximation from \cite{cora_semi-analytic_2018}:
\begin{equation}\label{eqn:vrps}
     r_{\mathrm{rps}} = -\tfrac{1}{2} \, r_d \, \ln \Big[ \rho_{\mathrm{gas}} \left(V_{\mathrm{gal}}^{\mathrm{tot}}\right)^2 \left(2 \pi G \, f_{\mathrm{gal}}^{\mathrm{gas}} \, \Sigma_0^2 \right)^{-1} \Big]
\end{equation}
where $\Sigma_0$ is the total surface density of each galaxy disk, $r_d$ the disk scale-length, and $f_{gal}^{gas}$ their respective gas fraction. The disk scale-length is taken from  $R_\mathrm{eff}=r_{1/2} = 1.68 r_d$, and $\Sigma_0$ from the masses displayed in
Table~\ref{tab:losvd_phot}. The gas fractions in galaxies are unknown, the density of the cluster hot gas $\rho_{gas}$ is
taken from  the previous considerations of the dark matter profile and from X-ray data \citep{2014Eckert,2017Eckert}.
To estimate the ram-pressure efficiency, the cold gas stripping radius has been computed according to equation \ref{eqn:vrps}.
Figure \ref{fig:RPS} plots this quantity for each galaxy as a function of the unknown gas fraction.

We also take into account that the gas stripping is maximum for a face-on inclination, and minimum for edge-on, as shown by \citet{Singh2024}, and have estimated $\varphi_{gal}$, the angle between the velocity vector and the spin vector for each galaxy.
\begin{figure}
    \centering
    \includegraphics[width=0.95\linewidth]{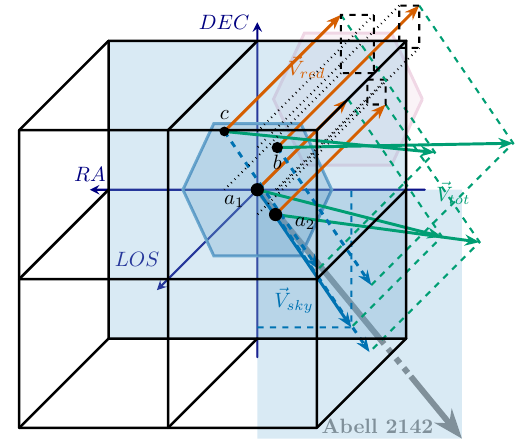}
    \caption{Kinematical configuration.  The four black bullets correspond to the position of the four galaxies on the sky.  
    3D velocities, labelled $V_{tot}$ and associated to each galaxy, are displayed as green arrows. Similarly, velocities in the plane of the sky, as derived from the X-ray tail, are displayed as blue arrows. The radial velocities associated to the redshift of each galaxy are displayed in red. The filled rectangles lie in the plane of the sky as well as the MANGA hexagonal field and are displayed in blue. (In order to give an overview of the 3D effect, we added a purple hexagon projected at the main redshift velocity.). 
    The gray arrow (in the plane of the sky) points towards the centre of A2142. 
    }
    \label{fig:vec}
\end{figure}
Relying on the direction of the X-ray tail, we estimate its position angle $\alpha=34^\circ$ with respect to the North direction. In the referential $(DEC,RA,LOS)$, we thus get the following coordinates for the velocity vector $\vv{V}(-V_{sky}\cos\alpha, -V_{sky}\sin\alpha, -V_{los})$. $V_{sky}$ is the global velocity modulus of the galaxies in the plane of the sky, assumed to be colinear with the direction of the X-ray tail. Fig.~\ref{fig:vec} displays schematically a 3D-rendering of the geometrical distribution of the group positions
and velocities. For the unit spin vector of a galaxy (PA, i), with the usual convention we get $\vv{s_{gal}}(- \sin i \cos\alpha, -\sin i \sin\alpha,- \cos i )$. We thus derive the $\varphi_{gal}$ angle between these two vectors as follows:
\begin{equation}
\cos \varphi_{\mathrm{gal}} =
\left(V_{\mathrm{sky}} / V_{\mathrm{tot}}\right)  \sin i
+ \left(V_{\mathrm{los}} / V_{\mathrm{tot}}\right) \cos i
\end{equation}
We thus get $\varphi_{gal}=57^\circ, 56^\circ, 57^\circ, 53^\circ$ for $a1$, $a2$, b and c, respectively, which introduces an uncertainty of about 1\,kpc on the r$_{rps}$ radius.

Regarding the gas content, we know that the galaxy group is actively forming stars with a global SFR of 45\,M$_\odot$\,yr$^{-1}$. Assuming the gas in a region with a typical radius of 5\,kpc with a standard depletion time ($\sim$ 2 Gyr), one expects $\Sigma_{SFR}=0.5 M_\odot$\,yr$^{-1}$\,kpc$^{-2}$. Following the typical 'Kennicutt-Schmidt' scaling relation  \citep[e.g. Fig. 3 in][]{2010MNRAS.407.2091G}, one can expect a molecular surface density of 
$\Sigma_{mol}=530$ M$_\odot$\,pc$^{-2}$. Assuming the gas concentrated within a radius of 5\,kpc, one can expect a total gas mass of $4.2\times 10^{10} f_{ff} M_\odot$, where $f_{ff}$ is the filling factor, which could be as low as 1\%. 

In comparison, the expected mass of the hot gas in the field of view (from the cluster) in a sphere of radius 13.5 kpc is $4 \times 10^7 M_\odot$. Similarly, one can estimate the mass of the cluster hot gas in the tail, supposed to be shaped as a cone (with 700\,kpc length and 150\,kpc half-width) is $ 10^{11} M_\odot$.

\subsection{Tidal effects}
\label{ssect:tidal}
The group of galaxies falling into the Abell 2142 cluster is quite compact, and its galaxies appear in obvious interactions The group contains four galaxies, two are merging (a1 and a2), and in the others tidal perturbations are detected as tidal tails and loops.
In order to provide an estimation of the degree of interaction, based on the estimated masses, 2D projected distances and relative radial velocities, we calculated the
tidal strength affecting the various galaxies. The total tidal strength produced by a neighbor M$_i$ on each galaxy, with a diameter d$_g$ and a mass M$_g$, is the sum of the tidal forces per unit mass, proportional to M$_i$d$_g$/D$_{ig}^3$, where  
M$_i$ is the mass of the neighbour, and D$_{ig}$ is its distance from
the centre of the considered galaxy. Approximating D$_{ig}$ by the projected separation, 
at the distance of the galaxy $g$. For each galaxy, we have summed on its three neighbors. We can then compare the total tidal force to the binding force of the galaxy $g$, i.e. proportional to M$_g$/d$_g^2$. The ratio of tidal to binding forces for each couple of galaxies is Q$_{ig}$ = M$_i$/M$_g$(d$_g$/D$_{ig}$)$^3$. The total interaction strength affecting a galaxy $g$ is the sum Q$_g$ = $\Sigma_i$Q$_{ig}$. This quantity has been estimated in Table \ref{tab:Tidal-Q}.

The most isolated galaxies identified in the sky have their ratio Q of about 10$^{-4}$, 
meaning that their total external forces amount only
to 0.01\% of their internal forces \citep{Verley2007}, while Hickson compact groups have Q between 0.1 to 30.
It results from this calculation that the group is highly interacting, with strength comparable
to the most compact of galaxy groups \citep{Hickson1992}. 

\begin{table}
    \centering
    \begin{tabular}{l|cccc}
    \hline
    Galaxy & a$_1$ & a$_2$ & b & c  \\
    \hline
     a$_1$  &      & 19.3 & 0.69 & 0.39 \\
     a$_2$  & 0.56 &      & 0.05 & 0.03 \\
     b      & 0.59 & 1.37 &      & 0.48 \\
     c      & 0.06 & 0.13 & 0.08 &      \\
     \hline
     Q$_g$  & 1.21 & 20.79& 0.82 & 0.89 \\
    \hline
    \end{tabular}
    \caption{Tidal strength, estimated first by Q$_{ig}$, then their sum Q$_g$.}
    \label{tab:Tidal-Q}
\end{table}


\section{Conclusion}
\label{sect:conclu}
We have studied the physical and dynamical state of the galaxy group infalling into the galaxy cluster Abell 2142, in order to better quantify the feedback effects of the cluster environment on galaxies. A large fraction of galaxy cluster growth is through such group accretion. This peculiar case is remarkable by the very long (700 kpc) ram pressure tail detected clearly in X-rays. 
The photometry of the system was made in ugr bands with $\sim$ 1 arcsec (1.7 kpc at z=0.0935) resolution with MegaCam on the CFH telescope, and integral field spectroscopy from MaNGA was used to derive stellar and gas kinematics, with 2.6 arcseconds (4.5 kpc), and 150-250 km/s resolution depending on wavelength. The full spectral energy distribution was used to derive the age and metallicity of the stellar populations using Nburst package. The gas kinematics and excitation are derived from the line emission of H$\alpha$, [NII], [OIII] and H$\beta$. The main difficulty is to separate the various components, corresponding to the different galaxies superposed on the line of sight. Given that only two spatial coordinates, and one radial velocity can be observed, the problem is not well determined, and the solution cannot be unique. For the sake of simplicity, we use parametric models of the four galaxies, with circular velocity and a variable dispersion, combined with simple mass models. We propose a solution compatible with the observations, within the error bars, with the four galaxy rotations disentangled. The kinematics is determined for both star and gas.  The galaxies are perturbed and intra-group gas is observed, certainly due to tidal stripping. Our model are disks dominated by rotation, although some regions reveal elevated dispersion, tracing the tidal perturbations.

The galaxy group is quite compact, and located at 1.3 Mpc from the cluster center. The group is so compact that it is difficult to disentangle tidal effects from ram-pressure stripping, while two of the galaxies are merging. The global star formation in the central four galaxies is estimated at 45 M$_\odot$/yr, corrected from AGN contribution, meaning that the galaxies are not yet quenched.  The galaxies are still rich in gas, and their ram pressure stripping is not yet extreme. They cannot be at the origin of the long X-ray tail, which must have come from the hot intra-group medium, which has been the first to be stripped at the galaxy group entry, infalling into the Abell 2142 cluster.

\subsection*{Acknowledgement} DG's research on the \nb\ fitting of the spectrum, and interpretation of the kinematic and properties of stellar population was supported by The Russian Science Foundation (RSCF) grants No.~23-12-00146.

\bibliographystyle{aa}
\bibliography{references,RAM}

\appendix

\section{Subaru images}
\label{app:subaru}
\begin{figure*}[!htb]
\centering
    \includegraphics[width=0.42\textwidth]{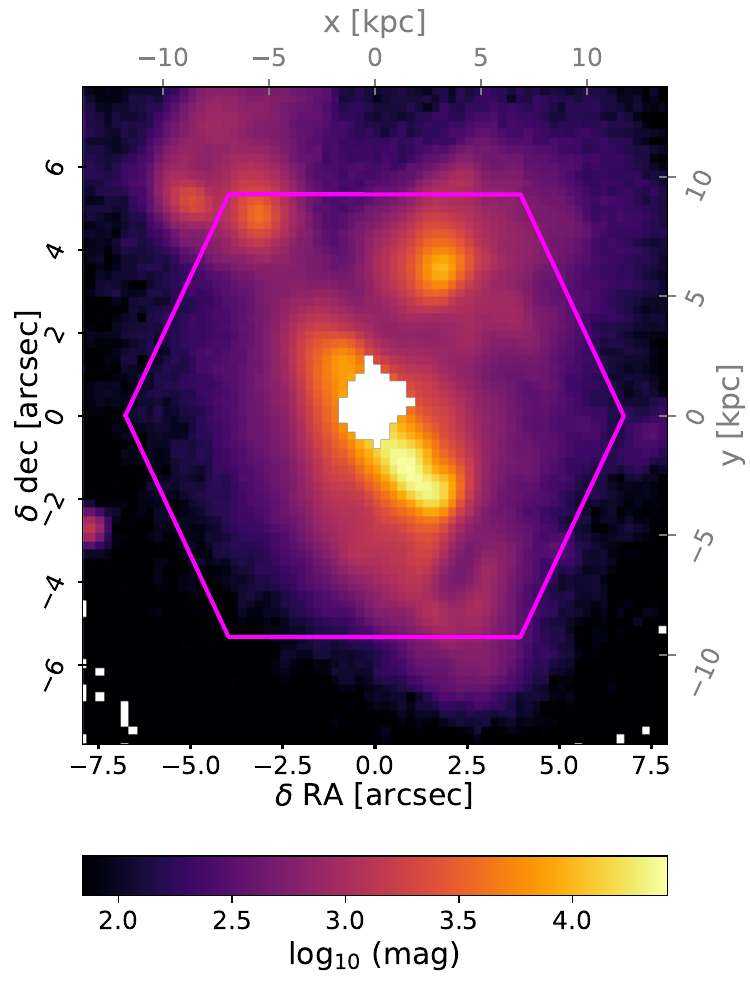}
    \includegraphics[width=0.42\textwidth]{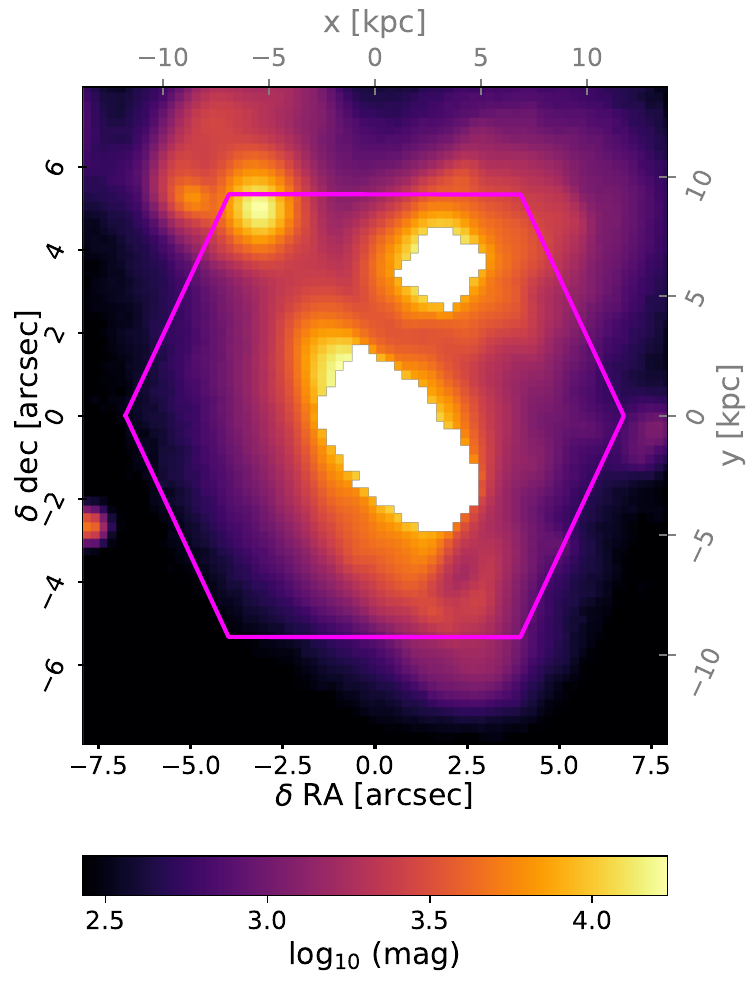}
    \caption{Subaru $g$-band (left) and $r$-band (right) images, from the JVO Subaru Suprime-Cam mosaic image archive. Available \href{https://jvo.nao.ac.jp/portal/subaru/spcam/dr2.do?action=mosaicInfo&mosaicId=SUPM19EA574B0393B100}{here for g} and \href{https://jvo.nao.ac.jp/portal/subaru/spcam/dr2.do?action=mosaicInfo&mosaicId=SUPM19E9E58C0472DE00}{here for r}.}
    \label{fig:subaru}
\end{figure*}

\section{Stellar Analysis}
\label{app:stellar-ana}

\begin{figure*}[!htb]
    \includegraphics[width=0.95\textwidth]{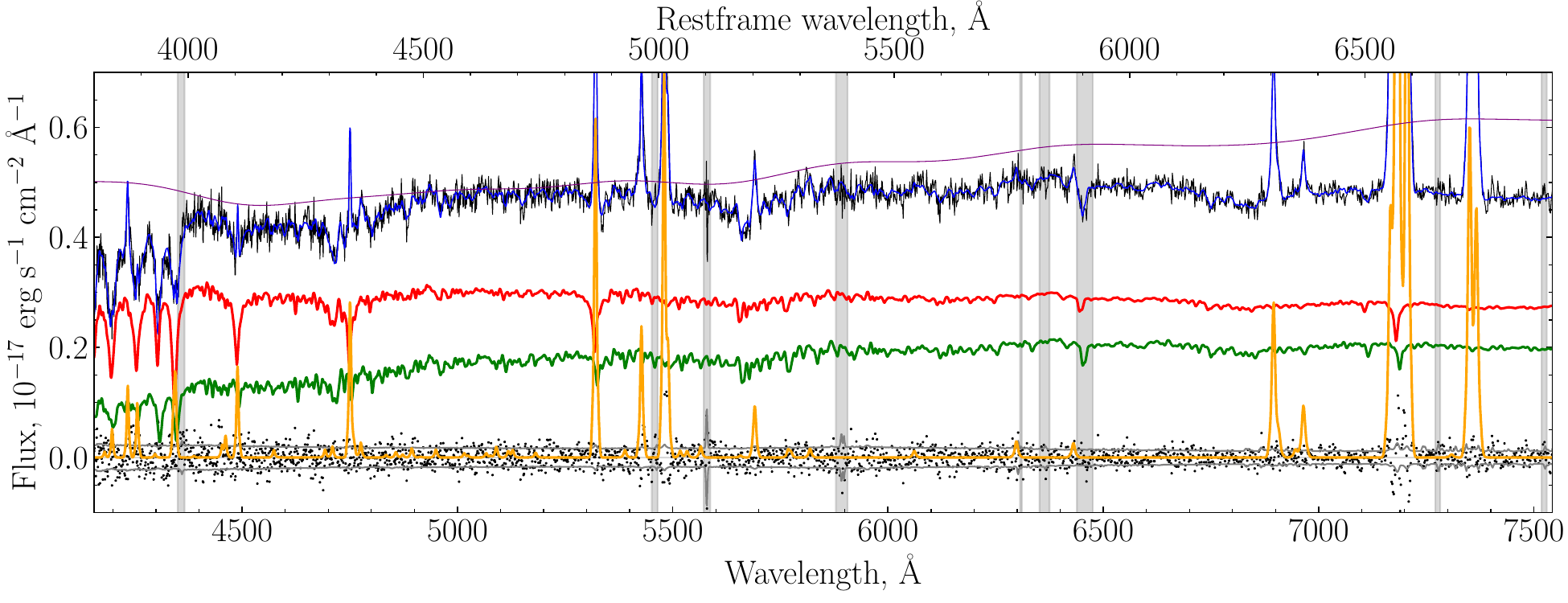}
    \caption{Example of 2-component \nb\ fitting for the spaxel with a$_1$ and b galaxies. Black line shows the observed spectrum in the spaxel and blue is the total best-fitting model. Red and green lines are stellar components of a$_1$ and b galaxies and the orange line is the emission lines component. Black dots are residuals and the gray line shows errors of spectrum. The shaded regions were excluded from fitting and the purple line is multiplicative continuum. The growing trend from blue to red spectrum part of the multiplicative continuum is a correction for the extinction.}
    \label{fig:2comp_fit}
\end{figure*}

\begin{figure*}[!htb]
    \centering \includegraphics[width=0.95\textwidth]{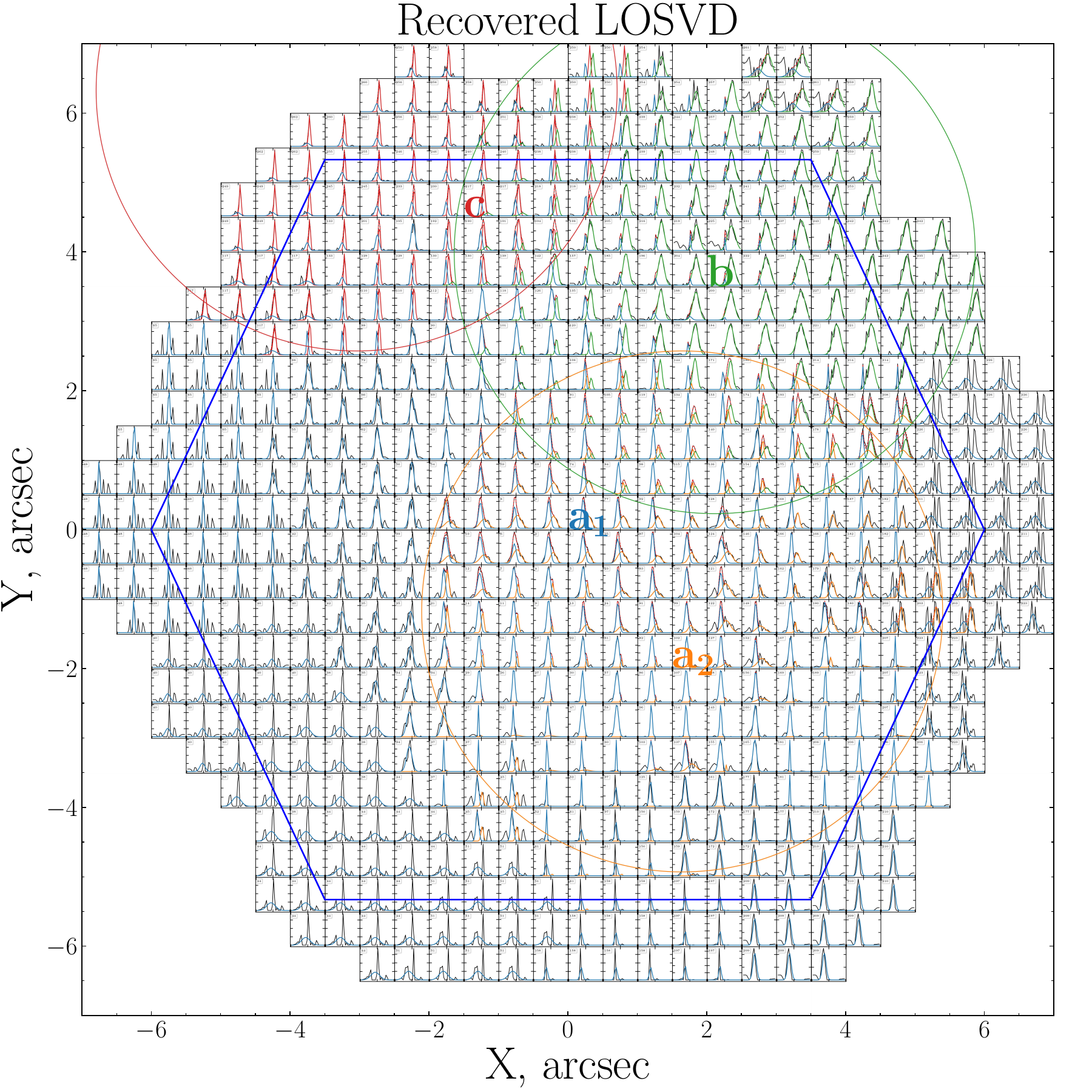}
    \caption{Non-parametric recovery of the stellar LOSVD. The panel shows the recovered LOSVD derived from the spectral binning in each spaxel of the MaNGA spectrum using Gaussian fitting techniques. The number of Gaussian components fitted varies according to the spatial location of each spaxel. Initially, all spaxels were fitted with a Gaussian representing the a$_1$ component (in blue). Additional Gaussian components were incorporated in regions where extra kinematic structures were identified (denoted by circles a$_2$, b, and c, and coded in orange, green and red).}
    \label{fig:LOSVD_bins}
\end{figure*}

\section{Gas analysis}
\label{broadcomp}

\begin{figure*}[!htb]
    \centering
    \includegraphics[width = 0.95\linewidth]{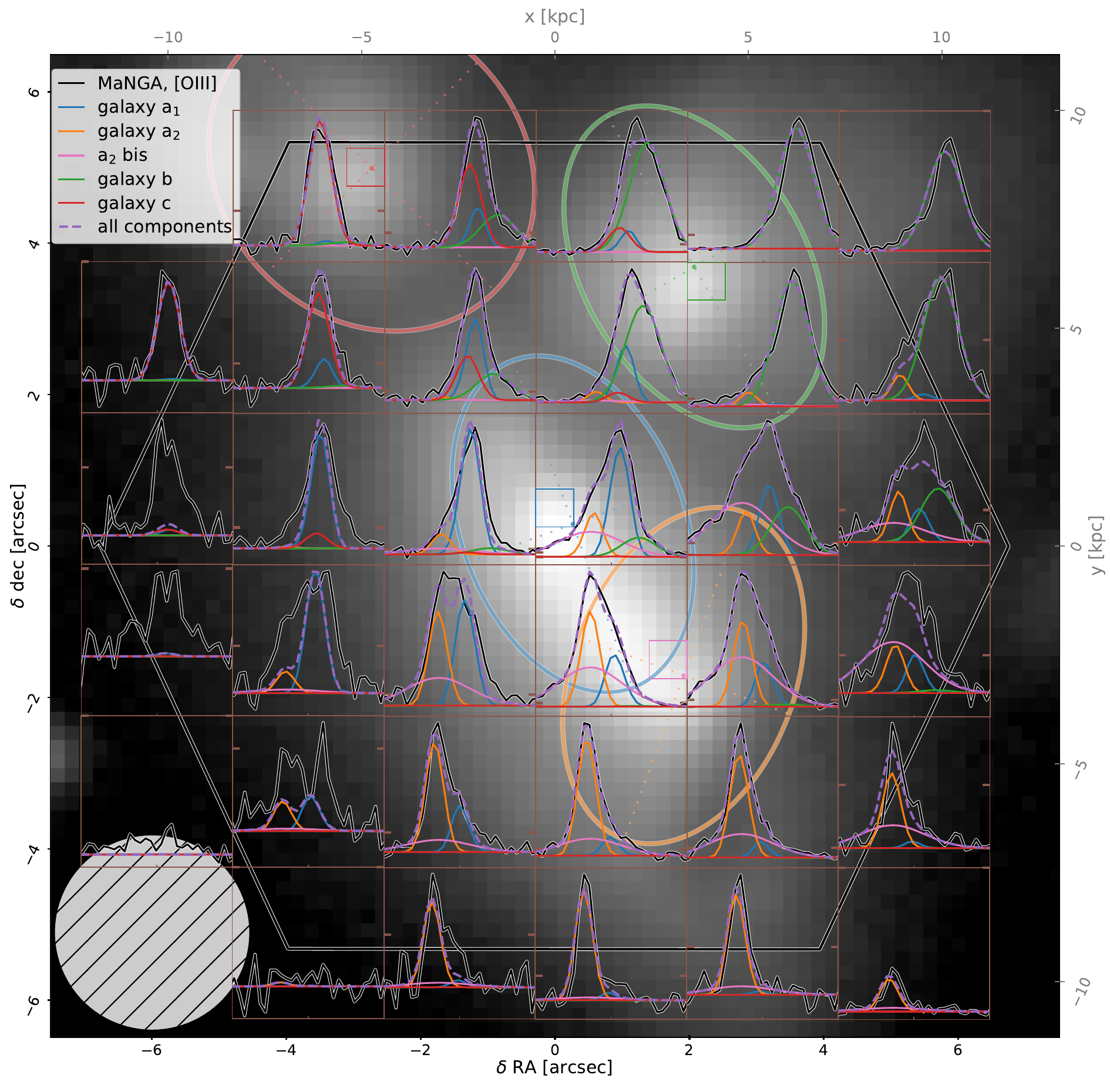}
    \caption{[OIII]~5008 Angstroms MaNGA observations and models of 4 galaxies and a broad component ($a_{\rm 2, bis}$), superimposed on the MegaCam $r$-band image. MaNGA pixels encompassing the centre of each modelled component are shown. Fits are done on original MaNGA spaxels but for the figure, MaNGA spectra and our fits are binned 16 by 16 and shown in brown boxes encompassing these binned spaxels. The velocity axis of each box is from -1000 to 1000 km/s. The two same ticks for values 0, on the baseline of spectra, and 0.01, are shown on the left and right $y$ axes of each box, to show the scale which is adapted to each box. The Gaussian reconstructed MaNGA PSF is shown on the bottom-left. The MaNGA hexagonal field-of-view is represented in black.} 
    \label{fig:spectra-oiii-fit}
\end{figure*}

\end{document}